\documentstyle[12pt,epsfig]{article}
\newcommand{\be}{\begin{equation}}
\newcommand{\ee}{\end{equation}}
\newcommand{\beq}{\begin{equation}}
\newcommand{\eeq}{\end{equation}}
\newcommand{\ba}{\begin{eqnarray}}
\newcommand{\ea}{\end{eqnarray}}
\newcommand{\bea}{\begin{eqnarray}}
\newcommand{\eea}{\end{eqnarray}}

\begin{document}
\baselineskip=15.5pt \pagestyle{plain} \setcounter{page}{1}
%--------+---------+---------+---------+---------+---------+---------+
%--------+---------+---------+---------+---------+---------+---------+
\begin{titlepage}

%\leftline{OUTP-10-yyyyy P}

%\vskip 0.8cm

\begin{center}

{\LARGE Plasma photoemission from string theory} \vskip .3cm

\vskip 1.cm

{\large {Babiker Hassanain\footnote{\tt babiker@thphys.ox.ac.uk}$
^{, \, a}$ and Martin Schvellinger\footnote{\tt
martin@fisica.unlp.edu.ar}$^{, \, b}$} }

\vskip 1.cm

{\it $^a$ The Rudolf Peierls Centre for Theoretical Physics, \\
Department of Physics, University of Oxford. \\ 1 Keble Road,
Oxford, OX1 3NP, UK.} \\

\vskip 0.5 cm

{\it $^b$ IFLP-CCT-La Plata, CONICET and \\
Departamento  de F\'{\i}sica, Universidad Nacional de La Plata.
\\ Calle 49 y 115, C.C. 67, (1900) La Plata,  \\ Buenos Aires,
Argentina.} \\

\vspace{1.cm}

{\bf Abstract}

\end{center}

Leading 't Hooft coupling corrections to the photoemission rate of
the planar limit of a strongly-coupled ${\cal {N}}=4$ SYM plasma are
investigated using the gauge/string duality. We consider the full
${\cal {O}}(\alpha'^3)$ type IIB string theory corrections to the
supergravity action, including higher order terms with the
Ramond-Ramond five-form field strength. We extend our previous
results presented in \cite{Hassanain:2011ce}. Photoemission rates
depend on the 't Hooft coupling, and their curves suggest an
interpolating behaviour from strong towards weak coupling regimes.
Their slopes at zero light-like momentum give the electrical
conductivity as a function of the 't Hooft coupling, in full
agreement with our previous results of \cite{Hassanain:2011fn}.
Furthermore, we also study the effect of corrections beyond the
large $N$ limit.

\noindent

\end{titlepage}

\newpage

\tableofcontents

\newpage

%\newpage
%--------+---------+---------+---------+---------+---------+---------+
%Body

\vfill

%%%%%%%%%%%%%%%%%%%%%%%%%%%%%%%%%%%%%%%%%%%%%%%%%%%%%%%%%%%%%%%%
\section{Introduction}
%%%%%%%%%%%%%%%%%%%%%%%%%%%%%%%%%%%%%%%%%%%%%%%%%%%%%%%%%%%%%%%%

Electrically charged particles in a quark gluon plasma (QGP) emit
photons. An analysis of these photons can lead to very valuable
information about the medium in which they are produced. On the one
hand, transport coefficients related to the electric charge such as
electrical conductivity and charge diffusion constant characterize
the dynamics of long wavelength, low frequency fluctuations in a
plasma. They are effectively related to ultra-soft photons, {\it
i.e.} those with momentum much smaller than the equilibrium
temperature of the medium, $T$. Ultra-soft photons eventually probe
the hydrodynamical regime of the plasma, with momentum $k \leq
\lambda^2 T$, where $\lambda$ is the 't Hooft coupling defined as
$\lambda \equiv g_{YM}^2 N$, where $g_{YM}$ is the SYM theory
coupling and $N$ the rank of its gauge group, $SU(N)$ in the present
case. On the other hand, it is possible to scrutinize a thermally
equilibrated plasma for a long range of emitted photon wavelengths.
This precisely gives shape to the photoemission rate from a plasma
as a function of the energy of the photons. It includes ultra-soft,
soft and hard photons, thus providing extremely useful information
about the dynamical structure of the medium.

For a weakly coupled QCD plasma, transport coefficients and
photoemission rates have been calculated using perturbative quantum
field theory in
\cite{Arnold:2000dr,Arnold:2001ba,Arnold:2001ms,Arnold:2002ja,
Arnold:2002zm,Arnold:2003zc} and references therein. These
references are particularly important since Arnold, Moore and Yaffe
have obtained the first complete leading order results for the
photoemission rates in QCD \cite{Arnold:2001ms}. They conclude that,
in addition to well known $2\longleftrightarrow 2$ particle processes, near-collinear {\it
Bremsstrahlung} and inelastic pair annihilation also make leading
order contributions. The Landau-Pomeranchuk-Migdal (LPM)
suppression, which is the effect produced by multiple soft
scatterings, may occur during the emission of the photon and has
important implications on the consistent treatment of the above
mentioned processes. The LPM effect leads to an ${\cal {O}}(1)$
suppression of these near-collinear processes.

There are indications, however, that the QGPs produced at the
Relativistic Heavy Ion Collider (RHIC) and at the Large Hadron
Collider (LHC) are in the strongly-coupled regime of QCD
\cite{Shuryak:2003xe,Gyulassy:2004zy,Muller:2007rs,
CasalderreySolana:2007zz,Shuryak:2008eq,Heinz:2008tv,Iancu:2008sp,
LHC-heavy-ion,Abreu:2007kv}. This is where the gauge/string
duality enters. This duality allows us to compute properties
of a strongly coupled gauge theory in terms of a weakly coupled
holographic dual string theory description
\cite{Maldacena:1997re,Gubser:1998bc,Witten:1998qj}. We ought to
admit that at present there is no complete or exact holographic
string theory dual model which accounts for all the relevant properties
of real QCD, not even in the planar limit of the gauge theory. For reasons which
shall be explained below, the holographic string theory dual model
which has been considered so far for these investigations is in fact
dual to the large $N$ limit of the strongly-coupled $SU(N)$ ${\cal
{N}}=4$ supersymmetric Yang-Mills (SYM) plasma.

The holographic dual model of the planar limit of the
strongly-coupled $SU(N)$ ${\cal {N}}=4$ SYM plasma is defined in
terms of a type IIB supergravity background given by a direct
product of an Anti-de Sitter-Schwarzschild black hole in five
dimensions ($AdS_5 BH$) times a five sphere $S^5$. There is a number
of considerations to take into account at the moment of
extrapolating this dual description of the large $N$ limit of ${\cal
{N}}=4$ SYM theory in order to make contact with QCD. Firstly, as it
is well known at zero temperature these theories are very different.
Indeed, the field content is different: while QCD has three colour
degrees of freedom and three flavours, matter is in the fundamental
representation of the gauge group $SU(3)$, it shows colour
confinement, has explicit and spontaneous chiral symmetry breaking,
and displays a discrete spectrum; on the other hand, in the ${\cal
{N}}=4$ SYM theory all their fields transform in the adjoint
representation of $SU(N)$, it is not a confining theory, conformal
symmetry is preserved at quantum level, and it is a supersymmetric
theory with the maximal number of supersymmetries in four
dimensions. At finite equilibrium temperature, $T$, above the
critical temperature of QCD, $T_c$, where hadrons become a
deconfined QGP, there are two regimes. For $T>>T_c$, again the two
types of plasmas related to these two theories behave very
differently too: while in QCD the coupling runs to weak coupling,
leading to a free gas of quarks and gluons; in the case of the
${\cal {N}}=4$ SYM plasma, the coupling, which remains constant, is
strong. Thus, it leads to a strongly coupled plasma. However, in the
intermediate region where $T$ is just above $T_c$, both plasmas
behave somewhat similarly. In this case QCD behaves as a strongly
coupled plasma of gluons and fundamental matter. These degrees of
freedom are deconfined, there is screening and the correlation
lengths are finite. Interestingly, the ${\cal {N}}=4$ SYM plasma
shares those properties because it is a strongly coupled plasma of
gluons and adjoint matter fields, it is also deconfined, shows
screening, and has finite correlation lengths. Moreover, quantum
field theories lattice calculations indicate that for certain
properties the similarities can be made even quantitatively (see for
instance \cite{CasalderreySolana:2011us} and references therein).
Therefore, one may assume that for $T>T_c$ but not $T>>T_c$, there
is a parametric region where one can focus on in order to describe
the rates for the emission of photons from a thermally equilibrated
SYM plasma using the gauge/string duality at finite yet strong 't
Hooft coupling.

A very important step towards the understanding of the photoemission
process and electric charge transport coefficients of QGP in terms
of the ${\cal {N}}=4$ SYM plasma has been done in a very nice paper
by Caron-Huot, Kovtun, Moore, Starinets and Yaffe
\cite{CaronHuot:2006te}. They consider the two limiting situations:
for very large and very small 't Hooft coupling. In the strongly
coupled case they consider the pure type IIB supergravity
description of the large $N$ limit of the ${\cal {N}}=4$ SYM plasma,
which we summarize in section 3. In the opposite limit they consider
a perturbative quantum field theory description of the ${\cal
{N}}=4$ SYM plasma, using similar ideas as in \cite{Arnold:2001ms}.
In section 5 we shall briefly review some perturbative results of
\cite{CaronHuot:2006te}.

In the light of the new experimental findings suggesting that the
QGP plasma at RHIC and LHC is in the strongly coupled regime of QCD,
a more realistic outlook requires a consideration of the 't Hooft
coupling expansion around the infinitely strongly-coupled regime of
the plasma. On the string theory side, we must therefore consider
the full ${\cal {O}}(\alpha'^3)$ type IIB string theory corrections
to the supergravity action. It includes a number of terms which
arise from the supersymmetric completion to the standard power-four
ten-dimensional Weyl-tensor. These new terms are constructed from a
rank-6 tensor which contains the Ramond-Ramond five-form field
strength. This is indeed a very complicated task from the technical
point of view. However, it is worth carrying out since it yields the
precise structure of the 't Hooft coupling corrections to the strong
coupling regime. Using this procedure we have obtained very
interesting results, which we briefly describe here and present in
full detail in section 4.

Our results show the following features. Firstly, the slopes of the
photoemission rates, which at zero light-like momentum give the
electrical conductivity as a function of the 't Hooft coupling, are
in full agreement with our previous results of
\cite{Hassanain:2011fn}: the electrical conductivity increases as
the 't Hooft coupling decreases. This concerns the hydrodynamic
regime of the plasma. Secondly, for higher momentum, the height of
the peaks decrease as the 't Hooft coupling increases ({\it i.e.} as
we approach the limit of infinite coupling), their maxima are
shifted towards the ultraviolet and the photoemission rate curves
cross downwards the limiting strongly coupled curve for momentum
around 3 times the equilibrium temperature. Another important
feature which comes from our results is that the number of emitted
photons increases as the 't Hooft coupling weakens. These features
show an interpolating trend from the supergravity calculation of the
strongly coupled gauge theory towards the perturbative quantum field
theory calculation in the weakly coupled ${\cal {N}}=4$
supersymmetric Yang-Mills plasma. In addition to describing these
effects in more detail in the general discussion and conclusions
section below, we will also consider the effect of including
non-planar perturbative $1/N$ corrections from higher derivative
terms in the type IIB action as well as non-perturbative
contributions due to D-instanton effects.

In section 2 we briefly describe generalities about the formalism
behind the calculation of plasma photoemission rates based on the
computation of two-point correlators of electromagnetic currents. A
review of strongly coupled ${\cal {N}}=4$ SYM plasma results
entirely obtained within pure type IIB supergravity, {\it i.e.} with
no string theory corrections, is presented in section 3. In section
4, which is the longest section of the paper, we introduce details
of the formalism and results from our calculations of the leading 't
Hooft coupling corrections to the photoemission rate of a
strongly-coupled ${\cal {N}}=4$ SYM plasma using the gauge/string
duality. Section 5 is devoted to a very brief review of results in
the weakly coupled regime. The material of this section is used in
the last section of the article in order to carry out a general
discussion of our results.

%%%%%%%%%%%%%%%%%%%%%%%%%%%%%%%%%%%%%%%%%%%%%%%%%%%%%%%%%%%%%%%%
\section{Derivation of photoemission rates in SYM plasmas}
%%%%%%%%%%%%%%%%%%%%%%%%%%%%%%%%%%%%%%%%%%%%%%%%%%%%%%%%%%%%%%%%

In this section we very briefly review the formalism needed in order
to derive the photoemission rate in plasmas from thermal field
theory. Since we expect to be able to compare our results with those
of reference \cite{CaronHuot:2006te}, when possible we mainly follow
its notation through this paper. Also, the same assumptions as in
\cite{CaronHuot:2006te} are considered here: the plasma is in
thermal equilibrium; we do not include prompt photons produced by
the initial scattering of partons from the colliding nuclei; and,
the electromagnetic coupling constant, $e$, is considered small
enough in order to ensure that photons are not to be re-scattered.

Consider the minimal coupling of a photon to the electromagnetic
current $J^{\small\textrm{em}}_\mu$ of the SYM plasma. Recall that
the $SU(N)$ ${\cal {N}}=4$ gauge supermultiplet is $\{{\cal
{A}}_\mu^a, \psi_p, \phi_{pq}\}$, where $a$ is the $SU(N)$ colour
index, $p$, $q = 1, \cdot \cdot \cdot, 4$, and all the fields
transform in the adjoint representation of the gauge group. They are
$SU(N)$ gauge bosons, 4 Weyl fermions and 6 real scalars,
respectively. Furthermore, since there is an anomaly free global
$SU(4)$ $R$-symmetry, there is an associated global $R$-symmetry
current, $J_\mu$. The way to consider the electromagnetic coupling
is by adding a $U(1)$ gauge field $A^\mu$ which couples to the
conserved current of a $U(1)$ subgroup of the full $SU(4)$
$R$-symmetry group \cite{CaronHuot:2006te}, under the assumption
that, to leading order in $e$,  $J^{\small\textrm{em}}_\mu \equiv
J_\mu$. Thus, the Lagrangian can be written as
\begin{equation}
L = L_{SYM} + e \, J^{\small\textrm{em}}_\mu \, A^\mu - \frac{1}{4}
F_{U(1)}^2 \, ,
\end{equation}
where $L_{SYM}$ is the Lagrangian of the ${\cal {N}}=4$ SYM theory
and $F_{U(1)}^2$ is the kinetic term of the photon field.

We denote the photon four-momentum as $K \equiv (k^0, \vec{k})$,
which is a null four-vector having its time component fixed by the
on-shell condition $k^0 = |\vec{k}|$. We use the mostly plus
signature for the Minkowski metric in four dimensions, denoted by
$\eta_{\mu\nu}=(-,+,+,+)$. First, let us consider the Wightman
function of electromagnetic currents defined as
\be
C_{\mu\nu}^<(K) = \int d^4X \, e^{-i K \cdot X} \,
<J_\mu^{\small\textrm{em}}(0) J^{\small\textrm{em}}_\nu(X)> \, ,
\ee
which in thermal equilibrium is related to the spectral density
$\chi_{\mu\nu}(K)$ by
\be
C_{\mu\nu}^<(K) = n_b(k^0) \, \chi_{\mu\nu}(K) \, ,
\ee
with the Bose-Einstein distribution function $n_b(k^0)=1/(e^{\beta
k^0}-1)$. In addition, the spectral density is given by the
imaginary part of the retarded current-current correlation function
\be
\chi_{\mu\nu}(K) = -2 \,  \textrm{Im} \,
C^{\small\textrm{ret}}_{\mu\nu}(K) \, .
\ee
The number of photons which are produced per unit time per unit
volume is denoted by $\Gamma_\gamma$. At leading order in $e$ the
photoemission rate is given by
\be
d\Gamma_\gamma = \frac{e^2}{2 |\vec{k}|} \, \eta^{\mu\nu}
\,C^<_{\mu\nu}(K)\mid_{k^0=|\vec k|} \, \frac{d^3k}{(2\pi)^3} \, .
\ee
Notice that this formula for the photoemission rate holds to leading
order in the electromagnetic coupling $e$. On the other hand, and
very importantly for our purposes, it is valid non-perturbatively in
all other interactions, {\it i.e.} the strong interaction
\cite{CaronHuot:2006te}.

It is worth mentioning that the slope of the photoemission rate in
the zero-frequency limit is proportional to the electrical
conductivity of the plasma, $\sigma$, which can also be determined
by the current-current correlator using the Kubo formula:
\begin{equation}
\sigma = \lim_{k^0 \rightarrow 0} \, \frac{e^2}{6 T} \,
\eta^{\mu\nu} \,C^<_{\mu\nu}(k^0, \vec k=0) \, .
\end{equation}
In the next section we describe the computation of the plasma
photoemission rate in infinitely strongly-coupled plasma.

%%%%%%%%%%%%%%%%%%%%%%%%%%%%%%%%%%%%%%%%%%%%%%%%%%%%%%%%%%%%%%%%
\section{Review of photoemission rates at strong 't Hooft coupling}
%%%%%%%%%%%%%%%%%%%%%%%%%%%%%%%%%%%%%%%%%%%%%%%%%%%%%%%%%%%%%%%%

The $AdS_5 BH \times S^5$ background, which is an exact solution of
type IIB supergravity, is given by
\be
ds^2 = \left(\frac{r_0}{R}\right)^2\frac{1}{u} \, \left(-f(u) \,
 dt^2 + d\vec{x}^2\right) + \frac{R^2}{4 u^2 f(u)} \,
 du^2 + R^2  \, d\Omega_5^2 \, ,\label{metric}
\ee
where $f(u)=1-u^2$, and $R$ is the radius of the $AdS_5$ and the
five-sphere. The $AdS$-boundary is at $u=0$ and the black hole
horizon is at $u=1$. For the $AdS_5$ coordinates we use indices $m=
\{(\mu=0, 1, 2, 3),5\}$. It is well known that this is the
holographic dual background to the large $N$ limit of the $SU(N)$
${\cal {N}}=4$ SYM theory at finite temperature $T$.

As mentioned, the purpose of the present work is to investigate the
${\mathcal{O}}(\lambda^{-3/2})$ 't Hooft coupling corrections to the
photoemission rate of a $SU(N)$ ${\cal {N}}=4$ SYM plasma produced
by the leading order $\alpha'^3$ corrections to the pure type IIB
supergravity calculation. In this section we briefly review some of
the calculations of \cite{CaronHuot:2006te}, which are applicable
for the $\lambda \to \infty$ limit. The idea is to obtain the
correlation functions of two $R$-symmetry currents using the methods
developed in references \cite{Son:2002sd,Policastro:2002se}.

The general form of the correlator at finite temperature is obtained
by taking into account rotation and gauge invariance:
\be
C_{\mu\nu}^{\small {\textrm{ret}}}(K) =
\Pi^{\small{\textrm{T}}}(k^0, k) \,
P^{\small{\textrm{T}}}_{\mu\nu}(K) + \Pi^{\small{\textrm{L}}}(k^0,
k) \, P^{\small{\textrm{L}}}_{\mu\nu}(K) \, ,
\ee
where the transverse and longitudinal projectors are defined such
that $P^{\small{\textrm{T}}}_{0 \mu}(K) = 0$,
$P^{\small{\textrm{T}}}_{ij}(K) = \delta_{ij}-k_i k_j/k^2$, and
$P^{\small{\textrm{L}}}_{\mu\nu}(K) =
P_{\mu\nu}(K)-P^{\small{\textrm{T}}}_{\mu\nu}(K)$, with $P_{\mu\nu}
= \eta_{\mu\nu}-K_\mu K_\nu/K^2$. We use the notation for the photon
light-like momentum defined in the previous section and $k=|\vec
k|$. The trace of the spectral function is
\be
\chi_\mu^\mu(k^0, k) =  - 4 \, \textrm{Im}\Pi^{\small
{\textrm{T}}}(k^0, k) - 2 \, \textrm{Im}\Pi^{\small
{\textrm{L}}}(k^0, k) \, .
\ee
For light-like momentum only $\Pi^{\small {\textrm{T}}}$
contributes. Therefore, it is the only relevant part for the
computation of the photoemission rate.

The gauge/string duality establishes a precise prescription to
compute a two-point correlator of conserved currents in a strongly
coupled SYM theory. The idea is the following: the insertion of an
operator of the SYM theory at the $AdS$-boundary induces a
fluctuation of a certain ten-dimensional background field.
Specifically, using the gauge/string duality prescription, a global
$U(1)$ symmetry current in the SYM theory couples to a $U(1)$ gauge
field in the bulk, $A_m$. From the SYM theory point of view the
$U(1)$ group is a subgroup of the $SU(4)$ $R$-symmetry group of the
${\cal {N}}=4$ SYM theory. Recall that the $SU(4)$ group is
isomorphic to the $SO(6)$ group, which obviously is the global
symmetry which generates rotations among the 6 real scalars of the
vector supermultiplet of the gauge theory. On the other hand, from
the supergravity side, the isometry group of the five sphere is
$SO(6)$. Thus, there is a $U(1)$ subgroup, which is related to
vector fluctuations of the metric, whose gauge field is precisely
the $A_\mu$ Abelian gauge field. Therefore, the point is to solve
the linearised equations of motion for the vector perturbations of
the metric. The definition of the two-form field strength is
$F_{mn}=\partial_m A_n-\partial_n A_m$. With the identification $E_i
\equiv F_{0i}$ one can write down the EOMs for the vector
fluctuation by splitting them into the transverse $(x, y)$, and
longitudinal $(z)$ components as follows:
\ba
&& E''_{x,y} - \frac{2 u}{f(u)} \, E'_{x,y} +
\frac{\varpi_0^2-\kappa_0^2 f(u)}{u f^2(u)} E_{x,y} = 0 \, ,
\\
&& E''_{z} - \frac{2 \varpi_0^2 u}{f(u)(\varpi_0^2 - \kappa_0^2
f(u))} \, \, E'_{z} + \frac{\varpi_0^2-\kappa_0^2 f(u)}{u f^2(u)}
E_{z} = 0 \, ,
\ea
where primes denote derivatives with respect to the radial
coordinate $u$, and one defines $\varpi_0 \equiv k^0/(2 \pi T)$ and
$\kappa_0 \equiv k/(2 \pi T)$. The solution of these EOMs have been
discussed in \cite{CaronHuot:2006te}, so here we just quote their
results in the following equations. First, notice that the
correlators are determined by the boundary term of the
five-dimensional on-shell Maxwell action
\be
S_B = \frac{N^2 T^2}{16} \lim_{u \rightarrow 0} \int \,
\frac{d^4K}{(2 \pi)^4} \, \left[ \frac{f(u)}{\kappa_0^2
f(u)-\varpi_0^2} E_z'(u, K) E_z(u,-K)-\frac{f(u)}{\varpi_0^2}
E'_{x,y}(u, K) \cdot  E_{x,y}(u, -K) \right]
\ee
and by applying the Lorentzian AdS/CFT prescription
\cite{Son:2002sd} it turns out that the transverse component which
is the only one actually needed for the computation of the
photoemission rate is given by \cite{CaronHuot:2006te}
\be
\Pi^T(k^0, k) = - \frac{N^2 T^2}{8} \, \lim_{u \rightarrow 0}
\frac{E'_x(u, K)}{E_x(u, K)} \, .
\ee
For light-like momenta there is an analytical solution to the EOM
above which can be written in terms of a hypergeometric function
\be
E_x(u) = (1-u)^{-i \varpi_0/2} \, (1+u)^{-\varpi_0/2} \,  {_2F_1}
\left(1-\frac{1}{2}(1+i)\varpi_0, \, -\frac{1}{2}(1+i)\varpi_0; \,
1-i \varpi_0; \, \frac{1}{2}(1-u) \right) \, .
\ee
Thus, the trace of the spectral function for light-like momenta is
\be
\chi^\mu_{\,\,\,\mu}(k^0=k) = \frac{N^2 T^2 \varpi_0}{8} \,
\mid{_2F_1} \left(1-\frac{1}{2}(1+i)\varpi_0, \,
1+\frac{1}{2}(1-i)\varpi_0; \, 1-i \varpi_0; \, -1 \right) \mid^{-2}
\, .
\ee
In addition, the electrical conductivity is given by
\be
\sigma = e^2 \, \frac{N^2 T}{16 \pi} \, ,
\ee
which has been obtained from the Kubo formula quoted in the previous
section.

Finally, the photoemission rate is given by
\be
\frac{d\Gamma_\gamma}{dk} = \frac{\alpha_{\small\textrm{em}}N^2
T^3}{16 \pi^2} \, \frac{(k/T)^2}{e^{k/T}-1} \, \mid {_2F_1}
\left(1-\frac{(1+i)k}{4\pi T}, \, 1+\frac{(1-i)k}{4 \pi T}; \,
1-\frac{i k}{2 \pi T}; \, -1 \right)\mid^{-2} \, ,
\ee
which holds in the large $N$ limit and for large $\lambda$ (where
the supergravity approximation is valid, $1 <<
 \lambda << N$), and is valid for the whole range of photon energies.

Now, we proceed to investigate the leading 't Hooft coupling
corrections to these expressions and analyse their physical
implications.

%%%%%%%%%%%%%%%%%%%%%%%%%%%%%%%%%%%%%%%%%%%%%%%%%%%%%%%%%%%%%%%%
\section{'t Hooft coupling corrections to photoemission rates}
%%%%%%%%%%%%%%%%%%%%%%%%%%%%%%%%%%%%%%%%%%%%%%%%%%%%%%%%%%%%%%%%

In this section we present the general corrections to type IIB
supergravity action at leading order in $\alpha'$. Firstly, in
subsection 4.1 we describe the formalism needed to account for
higher derivative corrections to the effective IIB action. Then, we
focus on ${\mathcal{O}}(\alpha'^3)$ string theory corrections and
develop the vector perturbations we need for the computation of the
current-current correlators. In subsection 4.2 we carry out the
computation of 't Hooft coupling corrections to photoemission rates,
whose results we show in subsection 4.3. Our results
concerning the effects of leading $1/N$ corrections and
non-perturbative instanton contributions are restricted to the
electrical conductivity of the plasma, and are presented in the
discussion and conclusions, in the last section of the paper.

%%%%%%%%%%%%%%%%%%%%%%%%%%%%%%%%%%%%%%%%%%%%%%%%%%%%%%%%%%%%%%%%%%%%%%%%%%%%%
\subsection{Higher derivative corrections to the effective IIB action and vector perturbations}
%%%%%%%%%%%%%%%%%%%%%%%%%%%%%%%%%%%%%%%%%%%%%%%%%%%%%%%%%%%%%%%%%%%%%%%%%%%%%

To begin with, we consider the leading type IIB string theory
corrections to the supergravity action $S_{IIB}^{SUGRA}$ which are
given in the term $S_{{\cal {R}}^4}^{3}$. The total action that we
shall consider is
\be
S_{IIB}=S_{IIB}^{SUGRA} + S_{{\cal {R}}^4}^{3} \, .
\ee
At the strong 't Hooft coupling limit the holographic dual model is
derived from type IIB supergravity, {\it i.e.} for $\alpha'
\rightarrow 0$. This contains the Einstein-Hilbert action coupled to
the dilaton and the Ramond-Ramond five-form field strength
\be
S_{IIB}^{SUGRA}=\frac{1}{2 \kappa_{10}^2}\int \, d^{10}x\,
\sqrt{-G}\left[R_{10}-\frac{1}{2}\left(\partial\phi
\right)^2-\frac{1}{4.5!}\left(F_5\right)^2 \right] \,
.\label{action-10D}
\ee
Effects of higher curvature terms which includes ${\cal
{O}}(\alpha'^3)$, perturbative $1/N$ corrections as well as
instanton corrections were considered in the presence
of D3-branes in type IIB string theory by Green and Stahn in
\cite{Green:2003an}. This reference proposes a
supersymmetric completion of the $C^4$ term, where $C$ is the
ten-dimensional Weyl tensor, leading to the following correction:
\be
S_{{\cal {R}}^4}^{3} = \frac{\alpha'^3 g_s^{3/2}}{32 \pi G} \int
d^{10}x \int d^{16}\theta \sqrt{-g} \, f^{(0,0)}(\tau, \bar{\tau})
[(\theta \Gamma^{mnp} \theta)(\theta \Gamma^{qrs} \theta){\cal
{R}}_{mnpqrs}]^4 + c.c. \, , \label{green-stahn}
\ee
where $\tau$ is the complex scalar field given by $\tau_1 + i \tau_2
\equiv a + i e^{-\phi}$, with $a$ being the axion and $e^\phi=g_s$
the string coupling. The function $f^{(0,0)}(\tau, \bar{\tau})$ is
the so-called modular form. The tensor ${\cal {R}}$ tensor is
defined in terms of the Weyl tensor and
\be
F^+=(1+*)F_5/2 \, ,
\ee
as given in
\cite{Green:2003an,deHaro:2002vk,Paulos:2008tn,Myers:2008yi}
\be
{\cal {R}}_{mnpqrs} = \frac{1}{8} g_{ps} C_{mnqr} + \frac{i}{48} D_m
F^+_{npqrs} + \frac{1}{384} F^+_{mnpkl} F_{qrs}^{+\,\,\, kl} \, .
\ee
The action (\ref{green-stahn}) was arrived at using the fact that the
physical content of type IIB supergravity can be arranged in a
scalar superfield $\Phi(x, \theta)$, where $\theta_a$, with $a = 1,
\cdot \cdot \cdot, 16$, is a complex Weyl spinor of $SO(1, 9)$. The
matrices $\Gamma$ have the usual definitions \cite{Paulos:2008tn}.

The modular form is presented in \cite{Green:1997tv} and is given by
the following expression
\be
f^{(0,0)}(\tau, \bar{\tau}) = 2 \zeta(3) \tau_2^{3/2} + \frac{2
\pi^2}{3} \tau_2^{-1/2} + 8 \pi \tau_2^{1/2} \sum_{m \neq 0, n \geq
0} \frac{|m|}{|n|} e^{2 \pi i |m n| \tau_1} K_1(2 \pi |m n| \tau_2)
\, ,
\ee
where $K_1$ is the modified Bessel function of second kind which
comes from the non-perturba\-ti\-ve D-instantons contributions.
Recall that the zeta function $\zeta(3)$ is the coefficient of the
first perturbative correction in the Eisenstein series of the
modular form. Note that in the background we consider with $N$
coincident parallel D3 branes, the axion vanishes, thus $\tau_1 =
0$, while $\tau_2 = g_s^{-1}$. Therefore, for small values of $g_s$
the modular form becomes
\be
f^{(0,0)}(\tau, \bar{\tau}) = 2 (4 \pi N)^{3/2} \left(
\frac{\zeta(3)}{\lambda^{3/2}} + \frac{\lambda^{1/2}}{48 N^2} +
\frac{e^{-8 \pi^2 N/\lambda}}{2 \pi^{1/2} N^{3/2}} \right) \, .
\ee
It is interesting to mention that Green and Stahn also have shown
that the D3-brane solution in supergravity does not get renormalised
by higher derivative terms \cite{Green:2003an}. Previously Banks and
Green had shown that $AdS_5 \times S^5$ is a solution to all orders
in $\alpha'$ \cite{Banks:1998nr}.

Now, we focus on the large $N$ limit of the dual $SU(N)$ ${\cal
{N}}=4$ SYM theory. Later on, in the conclusions, we shall return to
the consequences of the general corrections to the electrical
conductivity.

The finite leading 't Hooft coupling corrections are accounted
for by the following action \cite{Myers:2008yi}
\be\label{10DWeyl}
S_{IIB}^{\alpha'}=\frac{R^6}{2 \kappa_{10}^2}\int \, d^{10}x\,
\sqrt{-G}\left[ \, \gamma e^{-\frac{3}{2}\phi} \left(C^4 +
C^3{\mathcal{T}}+C^2{\mathcal{T}}^2+C{\mathcal{T}}^3+{\mathcal{T}}^4
\right)\right] \, , \label{10d-corrected-action}
\ee
obtained from the action (\ref{green-stahn}) in the large $N$ limit,
where $\gamma \equiv \frac{1}{8} \, \zeta(3)
\, (\alpha'/R^2)^{3}$, where $R^4 = 4 \pi g_s N \alpha'^2$. Since
$\lambda = g_{YM}^2 N \equiv 4 \pi g_s N$, we get $\gamma =
\frac{1}{8} \, \zeta(3) \, \frac{1}{\lambda^{3/2}}$. This action was
computed in \cite{Paulos:2008tn}, using the methods of
\cite{Green:2005qr}.

The $C^4$ term is a dimension-eight operator, defined as follows:
\be
C^4=C^{hmnk} \, C_{pmnq} \, C_h^{\,\,\,rsp} \, C^{q}_{\,\,\,rsk} +
\frac{1}{2} \, C^{hkmn} \, C_{pqmn} \, C_h^{\,\,\, rsp} \,
C^q_{\,\,\, rsk} \, ,
\ee
where $C^{q}_{\,\,\, rsk}$ is the Weyl tensor. The tensor ${\cal
{T}}$ is defined by
\begin{equation}
{\cal {T}}_{abcdef}= i\nabla_a
F^{+}_{bcdef}+\frac{1}{16}\left(F^{+}_{abcmn}F^{+}_{def}{}^{mn}-3
F^{+}_{abfmn}F^{+}_{dec}{}^{mn}\right) \, , \label{T-tensor}
\end{equation}
where the indices $[a,b,c]$ and $[d,e,f]$ are antisymmetrized in
each squared brackets, and symmetrized with respect to interchange
of $abc \leftrightarrow def$ \cite{Paulos:2008tn}.

At finite temperature the metric only gets corrections from the
$C^4$ term. This is so because the tensor ${\mathcal{T}}$ vanishes
on the uncorrected supergravity solution \cite{Myers:2008yi}. The
solution to the Einstein equations derived from the pure
supergravity action (\ref{action-10D}) is an $AdS_5BH \times S^5$
background. There are $N$ units of flux of $F_5$ through the sphere,
and the volume form of $S^5$ is denoted by $\epsilon$. On the field
theory side, $N$ is the rank of the gauge group, and it corresponds
to the number of parallel D3-branes whose back-reaction deforms the
space-time leading to the above metric in the near horizon limit.
The current operator $J_\mu(x)$ is dual to the $s$-wave mode of the
vectorial fluctuation on this background.

Next, we have to obtain the Lagrangian for the vectorial
perturbation in this background. Thus, we must construct a
consistent perturbed {\it Ansatz} for both the metric and the
Ramond-Ramond five-form field strength, such that a $U(1)$ subgroup
of the $SU(4)$ $R$-symmetry group is obtained
\cite{Argurio:1998cp,Cvetic:1999xp,Chamblin}. Then, by plugging this
consistent perturbation {\it Ansatz} into the full action (up to
${\mathcal{O}}(\alpha'^3)$) and integrating out the five-sphere, one
obtains the desired action for the $U(1)$ gauge field in the
$AdS_5BH$. Therefore, by studying the bulk solutions of the Maxwell
equations in the $AdS_5BH$ with certain boundary conditions we can
obtain the retarded correlation functions \cite{Son:2002sd,
Policastro:2002se,CaronHuot:2006te} of the operator $J_\mu(x)$.

Higher-curvature corrections to the type IIB supergravity action
correspond to finite 't Hooft coupling corrections in the field
theory. Suppose that we are interested in a certain observable of
the gauge theory, $\mathcal{O}$. If one carries out a series
expansion of it in inverse powers of the 't Hooft coupling one
schematically can write it as:
${\mathcal{O}}_0+{\mathcal{O}}_1/\lambda^{n_1}+\cdots$. The power
$n_1$ is a positive number corresponding to the leading $\alpha'^3$
correction to the type IIB supergravity action. In the present case,
we consider that ${\mathcal{O}}$ is the product of two
electromagnetic currents. Thus, we obtain the leading correction in
$\lambda$ using the gauge/string duality. The leading order
corrections come from terms ${\mathcal{O}}(\alpha'^3)$ in the
ten-dimensional action. It is important to recall that these
corrections dot no modify the metric at zero temperature
\cite{Banks:1998nr}. At finite temperature things are different as
shown in \cite{Gubser:1998nz,Pawelczyk:1998pb} where corrections to
the metric were obtained, and then further improved in
\cite{deHaro:2002vk,deHaro:2003zd,Peeters:2003pv}.

Higher curvature corrections on the spin-2 sector of the
fluctuations have been investigated in
\cite{Buchel:2004di,Buchel:2008sh,Hofman:2009ug,Sinha:2009ev}, among
other references. They are relevant to the computation of the
viscosity and mass-diffusion constants of the plasma.

In our case, we investigate vector fluctuations of the background.
The method to carry out the calculation consists of two steps.
Firstly, we have to obtain the minimal gauge-field kinetic term
using the vector-perturbed metric including the $\alpha'^3$
corrections to it, and the same for the five-form field strength.
Then, the corrections to the gauge field Lagrangian coming directly
from the higher-derivative operators have to be computed. The reason
why these two steps are different is that the first one will require
insertion of the corrected perturbation {\it Ans${\ddot{a}}$tze}
into the minimal ten-dimensional type IIB supergravity
two-derivative part Eq.(\ref{action-10D}). The second step requires
insertion of the {\it{uncorrected}} perturbation {\it
Ans${\ddot{a}}$tze} into the higher-curvature terms in ten
dimensions.

Our plan here is to start from the corrected metric and $F_5$
solutions, then proposing an {\it Ans${\ddot{a}}$tze} for the
perturbations that may be inserted into Eq.(\ref{action-10D}).

As mentioned before, the only piece of the ${\cal
{O}}(\alpha'^3)$-action which affects the metric is the $C^4$ term.
This induces the following corrected metric
\cite{Gubser:1998nz,Pawelczyk:1998pb,deHaro:2003zd}
\be
ds^2 = \left(\frac{r_0}{R}\right)^2\frac{1}{u} \, \left(-f(u) \,
K^2(u) \, dt^2 + d\vec{x}^2\right) + \frac{R^2}{4 u^2 f(u)} \,
P^2(u) \, du^2 + R^2 L^2(u) \, d\Omega_5^2 \, ,\label{proper-metric}
\ee
where we have used similar notation as for Eq.(\ref{metric}). The
functions of $u$ in the above metric are
\be
K(u) = \exp{[\gamma \, (a(u) + 4b(u))]} \, , \quad P(u) =
\exp{[\gamma \, b(u)]} \, , \quad L(u) =  \exp{[\gamma \, c(u)]} \,
,
\ee
where there are the following exponents, which are functions of the
radial coordinate
\ba
a(u) &=& -\frac{1625}{8} \, u^2 - 175 \, u^4 + \frac{10005}{16} \,
u^6 \, , \nonumber \\
b(u) &=& \frac{325}{8} \, u^2 + \frac{1075}{32} \, u^4
- \frac{4835}{32} \, u^6 \, , \nonumber \\
c(u) &=& \frac{15}{32} \, (1+u^2) \, u^4 \, .
\ea
In addition, the radius of the black hole horizon gets corrections
given by
\be
r_0 = \frac{\pi T R^2}{(1+\frac{265}{16} \gamma)} \, .
\ee
$T$ has been already identified as the physical equilibrium
temperature of the plasma. Thus, having obtained the corrected
metric Eq.(\ref{proper-metric}), we have to focus upon the
appropriate perturbation {\it Ans${\ddot{a}}$tze}. The vectorial
perturbation we are interested in enters the metric and the $F_5$
solution, in contrast to the metric tensor perturbations - needed
for mass-transport phenomena in the hydrodynamical regime of the
plasma - the latter only enter the metric {\it Ansatz}, but not the
$F_5$ {\it Ansatz}. This observation obviously makes the computation
of the corrections to the mass-transport coefficients much more
straightforward compared with the electric charge-transport
coefficients as well as other plasma properties beyond the
hydrodynamical domain.

We first obtain the kinetic term for the gauge fields. For this
purpose we plug the corrected {\it Ansatz} into the two-derivative
supergravity action Eq.(\ref{action-10D}). The metric {\it Ansatz}
reads
\ba
ds^2&=&\left[ g_{mn}+ \frac{4}{3} R^2 L(u)^2\, A_m A_n \right]
\,dx^m dx^n + R^2 L(u)^2 \, d\Omega_5^2  + \frac{4}{\sqrt{3}} R^2
L(u)^2 \nonumber \\
&& \times \left(\sin^2y_1 \, dy_3 + \cos^2 y_1 \, \sin^2 y_2 \, dy_4
+ \cos^2 y_1 \, \cos^2 y_2 \, dy_5 \right) \, A_m \, dx^m \,
\label{metric-ansatz} ,
\ea
where the metric of the unit five-sphere is given by
\be
d\Omega_5^2  = dy_1^2 + \cos^2 y_1 \, dy_2^2  + \sin^2 y_1 \, dy_3^2
+ \cos^2 y_1 \, \sin^2 y_2 \, dy_4^2  + \cos^2 y_1 \ \cos^2y_2 \,
dy_5^2 \, .\nonumber
\ee
Notice that since we are only interested in the terms which are
quadratic in the gauge-field perturbations we can write the $F_5$
{\it Ansatz} as follows
\be
G_5 = -\frac{4}{R} \overline{\epsilon} + \frac{R^3 L(u)^3}{\sqrt{3}}
\, \left( \sum_{i=1}^3 d\mu_i^2 \wedge d\phi_i \right) \wedge
\overline{\ast} F_2 \, , \label{F5ansatzCorrected}
\ee
where $F_2 = dA$ is the Abelian field strength and
$\overline{\epsilon}$ is a deformation of the volume form of the
metric of the $AdS_5$-Schwarzschild black hole. We should mention
that we are not interested in the part of $G_5$ which does not
contain the vector perturbations because it only contributes to the
potential of the metric, and is thus accounted for by the use of the
corrected metric in the computation. The Hodge dual $\ast$ is taken
with respect to the ten-dimensional metric, while $\overline{\ast}$
denotes the Hodge dual with respect to the five-dimensional metric
piece of the black hole. In addition, we have the usual definitions
for the coordinates on the $S^5$
\ba
&& \mu_1 = \sin y_1 \, ,\,\,\,\,\,\,\,\,\,\,\,\,\,\,\,\,\,\,\, \mu_2
= \cos y_1 \, \sin y_2 \, , \,\,\,\,\,\,\,\,\,\,\,\,\,\,\,\,\,\,\,
\mu_3 = \cos y_1 \, \cos y_2 \, ,  \nonumber \\
&& \phi_1 = y_3 \, ,
\,\,\,\,\,\,\,\,\,\,\,\,\,\,\,\,\,\,\,\,\,\,\,\,\, \phi_2 = y_4 \, ,
\,
\,\,\,\,\,\,\,\,\,\,\,\,\,\,\,\,\,\,\,\,\,\,\,\,\,\,\,\,\,\,\,\,\,\,\,\,\,\,\,\,\,\,
\phi_3 = y_5 \, .
\ea
By inserting these {\it Ans\"atze} into Eq.(\ref{action-10D}), and
discarding all the higher (massive) Kaluza-Klein harmonics of the
five-sphere, we get the following action for the zero-mode Abelian
gauge field $A_m$
\be
S_{IIB}^{SUGRA} = -\frac{{\tilde {N}}^2}{64 \pi^2 R} \int d^4x \, du
\, \sqrt{-g} \, L^7(u) \, g^{mp} \, g^{nq} \, F_{mn} \, F_{pq} \, .
\label{Fsquared}
\ee
Above we have written the Abelian field strength, defined as
$F_{mn}=\partial_m A_n -
\partial_n A_m$, the partial derivatives are $\partial_m =
\partial/\partial x^m$, while $x^m=(t, \vec{x}, u)$, with $t$ and
$\vec{x}=(x_1, x_2, x_3)$, are the Minkowski coordinates, and $g
\equiv \textrm{det} (g_{mn})$, which only involves the metric of
$AdS_5$-Schwarzschild black hole. Also notice that $L(u)$
straightforwardly comes from the dimensional reduction
\cite{Kovtun:2003wp}. The volume of the five-sphere has been
included in ${\tilde {N}}$.

Now, we should get the effect of the eight-derivative corrections of
Eq.(\ref{10DWeyl}). In order to achieve this we must determine the
five-dimensional operators that arise once the perturbed metric and
five-form field strength {\it Ans${\ddot{a}}$tze} are inserted into
Eq.(\ref{10DWeyl}). As in \cite{Hassanain:2010fv}, we use the
uncorrected {\it Ans${\ddot{a}}$tze} at this point. Indeed, we can
do it because using the corrected ones generates terms of even
higher order in $\gamma$. Clearly, the uncorrected {\it Ans\"atze}
are derived from the ones displayed here by taking
$L(u),K(u),P(u)\to 1$ and $\overline{\epsilon}\to \epsilon$. Next,
we explain how to calculate the explicit contributions from the
ten-dimensional operators, leading to the photoemission rates.

%%%%%%%%%%%%%%%%%%%%%%%%%%%%%%%%%%%%%%%%%%%%%%%%%%%%%%%%%%%%%%%%
\subsection{'t Hooft coupling corrections to photoemission rates}
%%%%%%%%%%%%%%%%%%%%%%%%%%%%%%%%%%%%%%%%%%%%%%%%%%%%%%%%%%%%%%%

In order to calculate the 't Hooft coupling corrections to
photoemission rates we now perform the explicit dimensional
reduction on $S^5$, including the leading type IIB string theory
corrections discussed in the previous subsection. This is done along
the lines of our previous work \cite{Hassanain:2011fn}\footnote{The
main difference with respect to our previous calculation of the
electrical conductivity of plasma in \cite{Hassanain:2011fn} is that
while for the conductivity it is only needed to consider the
dependence $A_m(u)$, for the photoemission rate it is necessary to
consider the dependence $A_m(t, z, u)$ which is not a trivial
extension of our former calculations in \cite{Hassanain:2011fn}.
Thus, having the $A_m(t, z, u)$ dependence implies actually a much
more complicated calculation.}. For this purpose it is necessary to
write explicitly all the terms of the full set of higher derivative
ten-dimensional operators which come from the supersymmetric
completion obtained in \cite{Green:2003an}. We use the definitions
introduced in \cite{Paulos:2008tn}
\be
C^4 +
C^3{\mathcal{T}}+C^2{\mathcal{T}}^2+C{\mathcal{T}}^3+{\mathcal{T}}^4
\equiv \frac{1}{86016} \sum_i n_i \, M_i \, .
\ee
Thus, we can write the two contributions to the $C^4$ term as
follows
\be
C^4 = - \frac{43008}{86016}  C_{abcd} C_{abef} C_{cegh} C_{dgfh} +
C_{abcd} C_{aecf} C_{bgeh} C_{dgfh} \, .
\ee
Repeated indices means usual Lorentz contractions. In order to
extract the quadratic terms in the vectorial fluctuations of the
metric we should notice that they can straightforwardly be computed
by expanding the ten-dimensional Weyl tensor as $C=C_0+C_1+C_2$,
where the sub-indices label the number of times that the Abelian
gauge field occurs. Obviously, a similar expansion can be made for
the ${\cal {T}}$ tensor: ${\cal {T}} = {\cal {T}}_0 + {\cal {T}}_1 +
{\cal {T}}_2$. In addition, from a straightforward explicit
calculation on the present background it can be shown that all the
components of ${\cal {T}}_2$ are zero. This fact is responsible of
an important simplification of the actual computations. Also, ${\cal
{T}}_0$ is zero for any compactification which contains a
five-dimensional Einstein manifold \cite{Myers:2008yi}, and
therefore it vanishes in the case we consider here.

Now, let us look at terms of the form $C^3 \, {\cal {T}}$;
\be
C^3 {\cal {T}} = \frac{3}{2} \, C_{abcd} C_{aefg} C_{bfhi} {\cal
{T}}_{cdeghi} \, .
\ee
Their only possible contributions comes in fact from terms like $C_1
C_0^2 {\cal {T}}_1$, $C_0 C_1 C_0 {\cal {T}}_1$ and $C_0^2 C_1 {\cal
{T}}_1$.

Then, let us study operators like $C^2 {\cal {T}}^2$. We find a few
contractions which can be collected in the following terms
\ba
C^2 {\cal {T}}^2 =  \frac{1}{86016} && \left( 30240 \, C_{abcd}
C_{abce} {\cal {T}}_{dfghij} {\cal {T}}_{efhgij} +  7392 \, C_{abcd}
C_{abef}
{\cal {T}}_{cdghij} {\cal {T}}_{efghij} \right. \nonumber \\
&& \left. - 4032 \, C_{abcd} C_{aecf} {\cal {T}}_{beghij} {\cal
{T}}_{dfghij} - 4032 \, C_{abcd} C_{aecf}
{\cal {T}}_{bghdij} {\cal {T}}_{eghfij} \right. \nonumber \\
&& \left. - 118272 \, C_{abcd} C_{aefg} {\cal {T}}_{bcehij} {\cal
{T}}_{dfhgij} -  26880 \, C_{abcd} C_{aefg}
{\cal {T}}_{bcehij} {\cal {T}}_{dhifgj} \right. \nonumber \\
&& \left. +  112896 \, C_{abcd} C_{aefg} {\cal {T}}_{bcfhij} {\cal
{T}}_{dehgij} - 96768 \, C_{abcd} C_{aefg} {\cal {T}}_{bcheij} {\cal
{T}}_{dfhgij} \right) \, . \nonumber \\
&&
\ea
The vanishing result of ${\cal {T}}_0$ implies that terms like $C_0
C_1 {\cal {T}}_0 {\cal {T}}_1$ also vanish. Then, the only possible
type of contribution from these terms is of the form $C_0^2 {\cal
{T}}_1^2$. Making use of the same arguments all the terms like $C
{\cal {T}}^3$ and ${\cal {T}}^4$ include a factor ${\cal {T}}_0$
and, therefore, are not present in a reduction upon a generic
five-dimensional Einstein manifold \cite{Hassanain:2010fv}.

Now, we proceed to explicitly calculate the operators above.
Firstly, we must calculate the ten-dimensional Weyl tensor with and
without vector fluctuations. Secondly, we need to obtain ${\cal
{T}}_1$, and by its definition it can be separated into one piece
which contains the covariant derivative, defined by
\be
({\nabla F_5})_{abcdef}= i\nabla_a F^{+}_{bcdef} \, ,
\ee
and a second piece which does not contain covariant derivatives
which reads
\be
\bar {\cal {T}}_{abcdef}  =
\frac{1}{16}\left(F^{+}_{abcmn}F^{+}_{def}{}^{mn}-3
F^{+}_{abfmn}F^{+}_{dec}{}^{mn}\right) \, .
\ee
So, we can write this tensor as ${\cal {T}}_1 = \nabla F_5 + {\bar
{\cal {T}}}$.

Let us define $F^{+} = F_{(e)} + F_{(m)}$. Thus, with the obvious
meaning of the electric and magnetic contribution, for the electric
part we have
\be
F_{(e)} = -\frac{4}{R} \epsilon + \frac{R^3}{\sqrt{3}} \, \left(
\sum_{i=1}^3 d\mu_i^2 \wedge d\phi_i \right) \wedge \overline{\ast}
F_2 \, ,
\ee
where $\overline{\ast}$ indicates the Hodge dual operation with
respect to the $AdS_5$-Schwarzschild black hole metric. It is
convenient to split the electric part into the background plus a
fluctuation,
\be
F_{(e)}=F_{(e)}^{(0)}+F_{(e)}^{(f)} \, ,
\ee
and similarly for the magnetic terms. Therefore, in components we
have
\be
(F_{(e)}^{(0)})_{\mu\nu\rho\sigma\delta} = - \frac{4}{R} \,
\sqrt{-g} \, \epsilon_{\mu\nu\rho\sigma\delta} \, ,
\ee
where $g$ is the determinant of the $AdS$ piece of the metric, in
fact $g=\det g_{AdS}= -r_0^4/(2 u^3 R^3)$. The Hodge dual gives
\be
(F_{(m)}^{(0)})_{abcde} =  - \frac{4}{R} \, R^5 \, \sqrt{\det{S^5}}
\epsilon_{abcde} \, .
\ee
Let us focus on the fluctuation. Actually, for this calculation we
only need the $U(1)$ gauge component $A_x(t, z, u)$, where in this
notation $t=x_1$ and $z=x_4$. Notice that if we were interested in
the electrical conductivity it is enough to consider the $A_x(u)$
dependence, which largely simplifies the calculation
\cite{Hassanain:2011fn} in comparison with the actual calculation of
the photoemission rates that we make in this work. Therefore, we
have to deal with the following non-vanishing components of the
two-form field strength: $F_{tx}$, $F_{zx}$ and $F_{ux}$, all of
them with the full dependence on $t, \, z$ and $u$
$AdS$-coordinates. We use the following definition:
$F=dA=\frac{1}{2!} F_{\mu\nu}/\sqrt{3} \, dx^\mu \wedge dx^\nu$.

So, the fluctuations of the electric part induce fluctuations in the
$F_5$ Ramond-Ramond field strength which are given by
\ba
(F_{(e)}^{(f)})_{y_iy_jtyz} &=& Fe_{ux}(t, z, u) \, b_{ij} \,
\epsilon_{y_iy_jtyz} \, , \\
(F_{(e)}^{(f)})_{y_iy_jyzu} &=& Fe_{tx}(t, z, u) \, b_{ij} \,
\epsilon_{y_iy_jyzu} \, , \\
(F_{(e)}^{(f)})_{y_iy_jtyu} &=& Fe_{zx}(t, z, u) \, b_{ij} \,
\epsilon_{y_iy_jtyu} \, ,
\ea
where
\ba
Fe_{ux}(t, z, u) &=& - \frac{R^3}{\sqrt{3}} \, \frac{1}{2} \,
\sqrt{-g} \, (2 F_{ux} G^{xx} G^{uu})
 \, , \\
Fe_{tx}(t, z, u) &=&  \frac{R^3}{\sqrt{3}} \, \frac{1}{2} \,
\sqrt{-g} \, (2 F_{tx} G^{tt} G^{xx})
 \, , \\
Fe_{zx}(t, z, u) &=&  \frac{R^3}{\sqrt{3}} \, \frac{1}{2} \,
\sqrt{-g} \, (2 F_{zx} G^{zz} G^{xx})
 \, ,
\ea
where the pairs $(i j)$ are $(1 3)$, $(1 4)$, $(1 5)$, $(2 4)$ and
$(2 5)$. The indices $i, j$ run over the coordinates of $S^5$, and
correspond to the coordinates $x_6, x_7, x_8, x_9$ and $x_{10}$. The
$b_{ij}$ functions are:
\ba
&& b_{13}=2 \sin y_1 \cos y_1 \, , \quad b_{14}= - 2 \sin^2 y_2 \sin
y_1 \cos y_1
\, , \quad b_{15}=- 2 \cos^2 y_2 \sin y_1 \cos y_1 \, , \nonumber \\
&& b_{24}= 2 \cos^2y_1 \sin y_2 \cos y_2 \, , \quad b_{25}= - 2
\cos^2 y_1 \sin y_2 \cos y_2 \, .
\ea
The fluctuations on the magnetic part are obtained after performing
the ten-dimensional Hodge dual operation on the corresponding
electrical fluctuations above. We present the full expression in the
appendix.

The kinetic term of the gauge field coming from the Ramond-Ramond
five-form field strength becomes
\be
-\frac{1}{4 \cdot 5!} \, F_5^2 = -\frac{2}{3} \, R^2 \, F^2 -
\frac{8}{R^2} \, ,
\ee
which is exactly what is expected. Recall that the scalar curvature
piece $R_{10}$ of the action gives $-1/3 \, R^2 \, F^2$, where $F^2$
denotes $F_{\mu\nu} F^{\mu\nu}$.

As we have seen in our previous paper \cite{Hassanain:2010fv}, the
eight-derivative ${\mathcal{O}}(\alpha'^3)$ corrections introduce a
large number of higher-derivative operators after the
compactification on a general five-dimensional Einstein manifold is
done. We must take account of them properly to solve the equation of
motion within perturbation theory. The situation is entirely
analogous to that studied in \cite{Buchel:2004di}, where the authors
were concerned with the tensor perturbations of the metric, but the
rationale is the same. We have discussed this for vectorial
perturbations of the metric in
\cite{Hassanain:2011fn,Hassanain:2010fv}. Lagrangian for the
transverse mode $A_x$ reads
\ba
S_{\textrm{total}} &=&- \frac{{\tilde {N}}^2 r_0^2}{16 \pi^2 R^4}
\int \frac{d^4k}{(2 \pi)^4} \int_0^1 du \,
\left[\gamma A_W A_k'' A_{-k}+ (B_1+\gamma B_W) A_k' A_{-k}' \right. \nonumber \\
&&\left. + \gamma C_W A_k' A_{-k}+(D_1+\gamma D_W) A_k A_{-k}
+\gamma E_W A_k'' A_{-k}''+\gamma F_W A_k'' A_{-k}' \right]  \, ,
\label{AxAction}
\ea
where we have introduced the following Fourier transform of the
field $A_x$,
\be
A_{x}(t, \vec{x}, u) = \int \frac{d^4k}{(2 \pi)^4} \, e^{-i \omega t
+ i q z} \, A_k(u) \, .
\ee
There are also a number of boundary terms that must be included for
this higher-derivative Lagrangian to make sense, and this is
discussed in detail in \cite{Buchel:2004di,Hassanain:2009xw}. The
coefficients $B_1$ and $D_1$ arise directly from the minimal kinetic
term $F^2$. The subscript $W$ indicates that the particular
coefficient comes directly from the eight-derivative corrections,
and the functions $A_W \to F_W$ are written below. Moreover, $B_1$
and $D_1$ contain some $\gamma$-dependence, but they are
non-vanishing in the $\gamma \to 0$ limit, while every other
coefficient vanishes in that limit. The equation of motion is given
by
\be
A_{x}''+p_1 A_{x}'+ p_0 A_{x} \, = \gamma \frac{1}{2 f(u)} G(A_{x})
\label{EqPotential} \, ,
\ee
where
\ba
G(A_{x})&=&A_W \, A_x''+C_W A_x'+2 \left(\delta D_1+D_W\right) A_x
-\partial_u\left(2 \delta B_1A_x'+2 B_W A_x'+C_W A_x +F_W A_x''
\right) \nonumber \\
&&+\partial_u^2 \left(A_W A_x+2 E_W A_x''+F_W
A_x'\right) \, ,
\ea
where  $B_1-B_1|_{\gamma\to 0}=\delta B_1$ and $D_1-D_1|_{\gamma\to
0}=\delta D_1$. First we have the coefficients with no
$\gamma$-dependence $p_0$ and $p_1$, given by
\be
p_0= \frac{\varpi_0^2-f(u)\kappa_0^2}{u f^2(u)} \quad \textrm{and}
\quad p_1=\frac{f'(u)}{f(u)} \, ,
\ee
where $\varpi_0=k_0/(2 \pi T)$ and $\kappa_0=k/(2 \pi T)$. For the
coefficients originating from the $F^2$ term in the action of the
gauge field, we obtain
\ba
B_1&=&\frac{K(u)f(u) L^7(u)}{P(u)} \, , \nonumber \\
D_1&=&-K(u)P(u) L^7(u)\left[\frac{\varpi^2-f(u)K^2(u)\kappa^2}{u
f(u)K^2(u)} \right] \, ,
\ea
where $\varpi=k_0 R^2/(2 r_0)$ and  $\kappa=k R^2/(2 r_0)$. At this
stage it is convenient to reduce the equation to a second-order
differential equation using a simple trick \cite{Cremonini:2009sy}.
The idea is that $\gamma A_x''=-\gamma \left(p_1 A_{x}'+ p_0
A_{x}\right)+{\mathcal{O}}(\gamma^2)$. Thus, we may reduce the
entire RHS of the equation of motion to terms which are first or
zeroth order in derivatives. The resulting equation is given by
\ba
&& A_{x}''+\left[ p_1-\frac{\gamma}{2 f(u)} \, [\theta_1(u)-p_1
\theta_2(u)] \right] A_{x}' + \left[p_0-\frac{\gamma}{2 f(u)} \,
[\theta_0(u)-p_0 \theta_2(u)] \right] A_{x} \, =
{\mathcal{O}}(\gamma^2) \, , \nonumber \\
&& \label{EOM}
\ea
where
\ba
&&\theta_0(u)=2 \, (\delta D_1 + D_W)- C_W' + A_W''
- 4 E_W' \, p_0' + 2 \, E_W \, (p_1 p_0'-p_0'') \, , \nonumber \\
&&\theta_1(u)=2 \, A_W'- 2 \, (\delta B_1+B_W)' + F_W''
- 4 E_W' \, (p_1'+p_0) + 2 E_W \, [p_1(p_1'+p_0) - p_1'' - 2 p_0'] \, , \nonumber \\
&&\theta_2(u)=2 \, A_W - 2 \, (\delta B_1+B_W) + F_W' + 2 E_W''-4
E_W' \, p_1 + 2 E_W \, [p_1^2-2 p_1'-p_0] \, . \nonumber \\
&&
\ea
In order to solve Eq.(\ref{EOM}), the first step is to examine the
singularity structure of the equation at the horizon $u=1$. As
usual, we change variables to $x=1-u$, so that the singularity is at
$x=0$, then insert the functional form $A_x=x^\beta$. We obtain the
indicial equation:
\be
\beta^2+\left(\frac{\omega}{4 \pi T}\right)^2=0 \,.
\ee
This is of course the same indicial equation that would have been
obtained in the infinite 't Hooft coupling limit. Thus, as long as
the Lagrangian originates from a gauge-invariant series of
operators, then the indicial equation is unchanged. The fact that
the indicial equation is unchanged is a consequence of the
gauge-invariance in five dimensions, which is in turn a consequence
of the $U(1)$ isometry of the internal manifold $S^5$, but it is not
a consequence of supersymmetry. This behaviour is expected
\cite{Buchel:2008sh,Ritz:2008kh,Myers:2009ij,Cremonini:2009sy} for
scalar and tensor fluctuations of the metric.

At this point, we have to solve the equation of motion for $A_x$.
First, we have to specify the functions $A_W, B_W, C_W, D_W, E_W$
and $F_W$ in the action Eq.(\ref{AxAction}). We have computed them
explicitly and obtained the following expressions:
\ba
A_W(t, z, u) &=& \frac{4 \gamma}{9} u^5 \left[157
\left(u^2-1\right) \kappa_0^2+275 \varpi_0^2\right]  \, , \\
B_W(t, z, u) &=& \frac{\gamma}{9} u^4 \left[u \left(-26214 u^3+29423
u+8844 \varpi_0^2+6012 \kappa_0^2 \left(u^2-1\right)\right)-9853\right] \, , \\
C_W(t, z, u) &=& \frac{4 \gamma}{9}  u^4 \left[\kappa_0^2 \left(3360
u^2-3046\right)-\frac{\left(1543 u^2+4540\right)
\varpi_0^2}{u^2-1}\right] \, , \\
D_W(t, z, u) &=& \frac{\gamma u^3}{9
   \left(u^2-1\right)^2}  \left[-3872 u \varpi_0^4+\left(1191 u^4+3857 u^2-5384 \kappa_0^2
\left(u^2-1\right) u+3796\right) \varpi_0^2 \right. \nonumber \\
&& \left. +\kappa_0^2
   \left(u^2-1\right)^2 \left(-1512 u \kappa_0^2+5241 u^2-2332\right)\right] \, , \\
E_W(t, z, u) &=& -\frac{3872 \gamma}{9}  u^6 \left[u^2-1\right]^2 \,
, \\
F_W(t, z, u) &=& -\frac{2 \gamma}{9}  u^5 \left(u^2-1\right)
\left[9719 u^2-6397\right] \, .
\ea

Finally, to enable us to carry out analytic computations of the
photoemission spectrum in the high-momentum limit, and as an aide to
the understanding of the physics behind the corrected equation of
motion, we can rewrite the EOM in the Schr\"odinger basis. We
transform the field variable as follows:
\ba
A_x(u)&=& \Sigma(u) \Psi(u) \, ,\nonumber \\
\Sigma(u)&=&\frac{1}{288} \sqrt{f(u)} \left(u^2 \gamma
\left(u^2 \left(37760 \kappa_0^2 u-87539 u^2+343897\right)-11700\right)+288\right)
\, . \label{transform}
\ea
giving us the Schr\"odinger-like equation:
\ba
\Psi''(u)&=&V(u)\Psi(u) \, \nonumber \\
V(u)&=&-\frac{1}{144 \left(u^2-1\right)^2} \left[144 (u
\kappa_0^2+1)+ \gamma (u^2-1) \left(1838319 u^6-4752055 u^4 \right.
\right.
\nonumber \\
&& \left. \left. +2098482 u^2+\kappa_0^2 \left(1011173 u^4+245442
   u^2-16470\right) u-11700\right)  \right]\, . \label{SPotential}
\ea
We can study the structure of the relative difference between this
$\gamma$-corrected potential and the uncorrected one (for $\gamma=0$
which corresponds to 't Hooft coupling going to infinity). Figure 1
shows the difference between the numerical potential minus the
analytical one for $\lambda \rightarrow \infty$, divided by the
analytical one for large $\lambda \rightarrow \infty$, as a function
of the radial coordinate of the black hole, $u$. Long-dashed,
dashed, small-dashed, tiny-dashed, and dotted lines correspond to
decreasing values of $\lambda=200, 150, 100, 50 $ and 35,
respectively. Notice the smooth behaviour of the corrected potential
as a function of $\lambda$ and also the fact that they coincide
exactly at $u=1$. This is behind the fact that the indicial equation
is unchanged from the uncorrected case. Notice also that there is an
exact cancelation of all $\kappa^4$ as well as $\kappa_0^4$ terms in
the potential, which is consequence of the supersymmetric structure
of the higher derivative corrections in the ten-dimensional
action\footnote{Notice that although in Eq.(\ref{SPotential}) we
have set the light-like momenta condition, the validity of the
statement about the power four-momentum and frequency terms
cancelation is quite general, and we have explicitly checked that.}.
As a consequence, the plasma structure functions show a slight
enhancement (at ultra-high momenta) from their values at $\lambda
\rightarrow \infty$ as the coupling decreases, as has been shown in
our previous work \cite{Hassanain:2009xw}.
\begin{figure}
\begin{center}
\epsfig{file=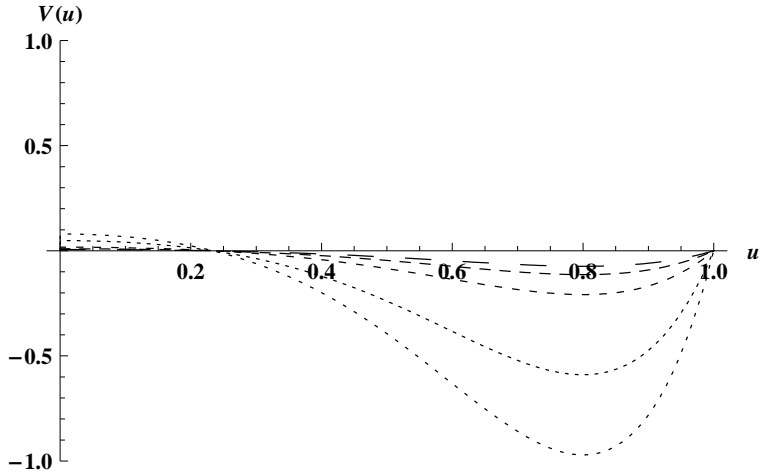, width=10cm} {\caption{\small Difference
between the numerical potential minus the analytical one for
$\lambda \rightarrow \infty$, divided by the analytical one for
$\lambda \rightarrow \infty$, as a function of the radial coordinate
of the black hole $u$. Long-dashed, dashed, small-dashed,
tiny-dashed, and dotted lines correspond to decreasing values of
$\lambda=200, 150, 100, 50 $ and 35, respectively.}}
\label{Schrodinger-Potential}
\end{center}
\end{figure}
The next step is to actually solve the EOM for $A_x$. We do this by
a numerical solution of Eq.(\ref{SPotential}). With this numerical
solution we compute the trace of the spectral function and the
photoemission rate for any value of the 't Hooft coupling at finite
yet strong coupling. We show our results in the next subsection.

%%%%%%%%%%%%%%%%%%%%%%%%%%%%%%%%%%%%%%%%%%%%%%%%%%%%%%%%%%%%%%%%%%%%%%%%%%%%%%%%%%
\subsection{Results of photoemission with 't Hooft coupling corrections }
%%%%%%%%%%%%%%%%%%%%%%%%%%%%%%%%%%%%%%%%%%%%%%%%%%%%%%%%%%%%%%%%%%%%%%%%%%%%%%%%%%

First, let us describe our results for the trace of the spectral
function. Its asymptotic behaviour can be evaluated analytically for
low- and high-momentum, and numerically for the remaining momentum
domain. This gives \cite{Hassanain:2011ce,CaronHuot:2006te}
\begin{equation}\label{chi-limit}
\frac{\chi^{\mu}_{\mu}(\kappa_0)}{\frac{1}{2}N^2T^2} = \left\{
\begin{array}{cc}
\left(1+\frac{14993}{9}\gamma \right) \kappa_0 +{\cal{O}}(k^3) & \, \kappa_0 \ll 1 \\
\frac{3^{5/6}}{2}
\frac{\Gamma\left(\frac{2}{3}\right)}{\Gamma\left(\frac{1}{3}\right)}\left(1
+5\gamma\right)\kappa_0^{2/3}+{\cal{O}}(1)& \, \kappa_0 \gg 1
\end{array} \right. \,.
\end{equation}
The coefficient of $\kappa_0$ in the low-momentum regime of
Eq.(\ref{chi-limit}) means that the electrical conductivity of the
strongly-coupled plasma is enhanced by a factor
$\left(1+\frac{14993}{9}\gamma \right)$ due to the finite $\lambda$
corrections \cite{Hassanain:2010fv}. This is as expected from the
perturbative computations in \cite{CaronHuot:2006te} since the
weakly-coupled plasma has a larger mean-free-path per collision,
allowing more efficient diffusion of electric charge, and hence a
higher electrical conductivity. On other hand, for the higher
momentum, the results of \cite{CaronHuot:2006te} imply that the
spectral function at weak coupling should go like $\kappa_0^{1/2}$
in the ultraviolet. Given the fact that the spectral function at
$\lambda \rightarrow \infty$ goes like $\kappa_0^{2/3}$, in that
regime one would have expected our result in Eq.(\ref{chi-limit}) to
display some smooth interpolation between $\kappa_0^{1/2}$ and
$\kappa_0^{2/3}$. We do not obtain such an interpolation, finding
instead that the finite coupling corrections do not change the
momentum-dependence for large momentum. Moreover, we find an
enhancement by a factor $(1+5\gamma)$ in that regime (see also
\cite{Hassanain:2009xw}). The fact that the leading $\kappa_0$
behaviour is unchanged by the corrections could have been seen from
the Schr\"odinger-like potential above: the only
$\kappa_0$-dependence is $\kappa_0^2$, identically to the $\lambda
\rightarrow \infty$ case. Terms like $\kappa_0^4$, which could have
changed the high-momentum functional dependence of
$\chi^\mu_\mu(\kappa_0)$, vanish. Figure 2 shows the trace of the
spectral function $\chi^\mu_{\,\,\,\mu}$ divided by $\kappa_0$ as a
function of the light-like momentum $k$ for the cases when $\lambda$
is large. Solid, long-dashed, dashed and small-dashed lines
correspond to decreasing values of $\lambda=\infty, 200, 150$, and
100, respectively.
\begin{figure}
\begin{center}
\epsfig{file=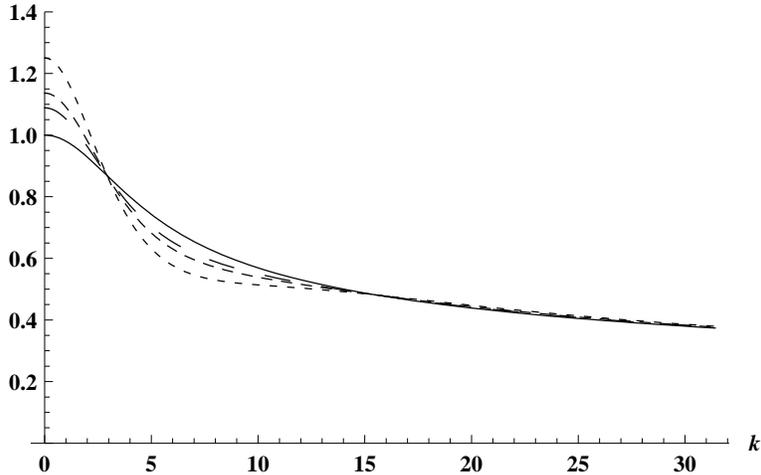, width=10cm} {\caption{\small Trace of the
spectral function $\chi_\mu^\mu$ divided by $\kappa_0$ as a function
of $k$ for the cases when $\lambda$ is large. Solid, long-dashed,
dashed and small-dashed lines correspond to decreasing values of
$\lambda=\infty, 200, 150$, and 100, respectively.}}
\label{Scpectral}
\end{center}
\end{figure}
In figure \ref{PhotoemissionRates} we show the photoemission rates
of a strongly-coupled ${\cal {N}}=4$ supersymmetric Yang-Mills
plasma as a function of photon momentum divided by the equilibrium
temperature, $k/T$. In fact the curves show $d\Gamma_\gamma/dk$
divided by $\alpha_{\small\textrm{em}} (N^2-1) T^3$. Different
curves correspond to different large values of the 't Hooft
coupling: solid, long-dashed, dashed, small-dashed, tiny-dashed, and
dotted lines correspond to decreasing values of $\lambda$ from
$\lambda=\infty$ (in fact it is the analytical expression from
supergravity with no string theory corrections), and then $\lambda=
200, 150, 100, 50$  and 35, respectively. These curves have been
obtained using the gauge/strings duality, considering the full
${\cal {O}}(\alpha'^3)$ type IIB string theory corrections to the
supergravity action. It is evident that the behaviour of the
photoemission rates depend upon the values of the 't Hooft coupling.
Their slopes at zero momentum give the electrical conductivity as a
function of the 't Hooft coupling in full agreement with our
previous results of \cite{Hassanain:2011fn}. The height of the peaks
decrease as the 't Hooft coupling increases, their maxima are
shifted towards the ultraviolet and the photoemission rate curves
cross downwards the limiting strongly coupled (pure supergravity)
curve for momentum around three times the equilibrium temperature.
These features are expected from perturbative quantum field theory
calculations in the weakly coupled ${\cal {N}}=4$ supersymmetric
Yang-Mills plasma and from the supergravity calculation of the large
$N$ strongly coupled theory.
\begin{figure}
\begin{center}
\epsfig{file=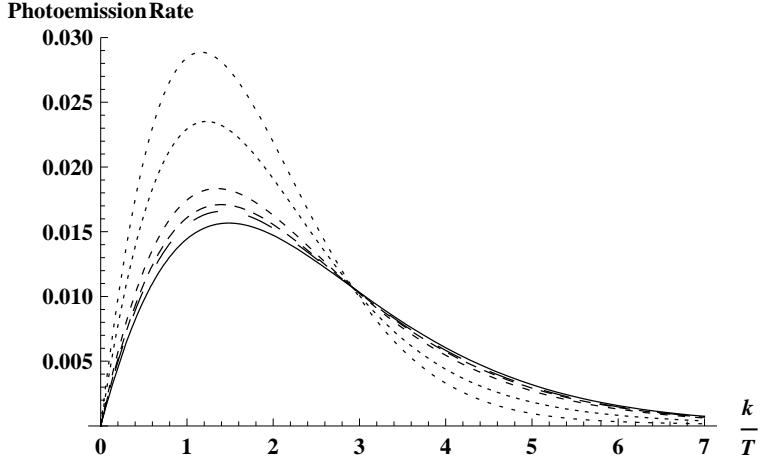, width=10cm} \caption{\small
Photoemission spectrum for different values of $\lambda$, as a
function of the light-like momentum of the emitted photon divided by
the equilibrium temperature, $k/T$. Notice that in fact we show the
curves for $d\Gamma_\gamma/dk$ divided by $\alpha_{\small
\textrm{em}} (N^2-1) T^3$. Solid, long-dashed, dashed, small-dashed,
tiny-dashed, and dotted lines correspond to decreasing values of
$\lambda$, from $\lambda=\infty$  (in fact it is the analytical
expression from supergravity with no string theory corrections), and
$200, 150, 100, 50 $ and 35, respectively.}
\label{PhotoemissionRates}
\end{center}
\end{figure}
However, at much higher momentum, all these curves cross upwards the
extreme strongly coupled one. This behaviour is exemplified in
figure \ref{PhotoemissionRatesLargek} for the large 't Hooft
coupling case (solid line) compared with $\lambda = 50$ (tiny-dashed
line), for a relatively large photon frequency in comparison with
the equilibrium temperature, actually the crossing occurs around $k
\approx 17.4 T$. For larger values of the momentum all curves
approach the solid one. This result is in agreement with our former
results on deep inelastic scattering structure functions from ${\cal
{N}}=4$ supersymmetric Yang-Mills plasma with string theory
corrections \cite{Hassanain:2009xw}. The reason for such a behaviour
comes from the fact that at ${\cal {O}}(\alpha'^3)$ in string
theory, the Schr\"odinger-like potential describing the dynamics of
the photo-production gets no corrections like the fourth power of
momentum and frequency. This is caused by an exact cancelation of
this power of momentum contributions from the $C^4$-term and its
supersymmetric completion at ${\cal {O}}(\alpha'^3)$ in the string
theory type IIB action.
\begin{figure}
\begin{center}
\epsfig{file=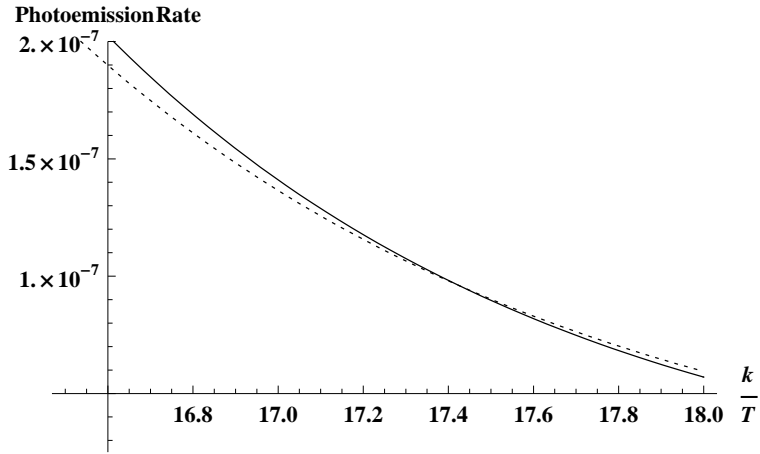, width=10cm}
\caption{\small Photoemission spectrum as a function of the
light-like momentum of the emitted photon divided by the equilibrium
temperature, $k/T$. As before we show the curves for
$d\Gamma_\gamma/dk$ divided by $\alpha_{\small \textrm{em}} (N^2-1)
T^3$. Solid and tiny-dashed lines correspond to  values of
$\lambda=\infty$ and $50 $, respectively. This figure shows the
typical large photon momentum behaviour having a crossing around
$k/T \approx 17.4$, for larger momentum the tiny-dashed curve
approaches the solid line.} \label{PhotoemissionRatesLargek}
\end{center}
\end{figure}

%%%%%%%%%%%%%%%%%%%%%%%%%%%%%%%%%%%%%%%%%%%%%%%%%%%%%%%%%%%%%%%%
\section{Weakly coupled SYM plasma photoemission rates}
%%%%%%%%%%%%%%%%%%%%%%%%%%%%%%%%%%%%%%%%%%%%%%%%%%%%%%%%%%%%%%%%

In this section we very briefly describe the results for the weakly
coupled regime of SYM obtained by \cite{CaronHuot:2006te}. We
include this in order to be able to compare our results in the
previous section for large values of $\lambda$ and see how they help
to understand a broader picture of the plasma structure in terms of
the 't Hooft coupling.

The computations of the spectral function in the weakly coupled
regime has been done using perturbative SYM theory
\cite{CaronHuot:2006te}, also previous results for perturbative QCD
were obtained in \cite{Arnold:2001ms,Arnold:2001ba}. For light-like
momenta the contribution to the trace of the spectral function
appears at two-loop level. The key point is that in a thermal system
the expansion of physical quantities in powers of the 't Hooft
coupling is not the same as the diagrammatic expansion in loops.
Basically, what happens is that there is sensitivity to energy and
momentum scales which are parametrically small in comparison with
the equilibrium temperature. In the situation where one considers
light-like momentum this complication already arises at the first
non-trivial order. To deal with this, it is necessary to carry out
an infinite resummation of diagrams in order to find the leading
order weak-coupling photon production rate. This rate can be split
into a contribution from a Compton-like $2 \leftrightarrow 2$
scattering process and near-collinear {\it Bremsstrahlung} and
pair-annihilation processes, which are further corrected due to the
Lipatov-Pomeranchuk-Migdal suppression effect. The complete
presentation of all these processes and effects for QCD is given in
\cite{Arnold:2001ms,Arnold:2001ba} and references therein. Also in
\cite{CaronHuot:2006te} is presented a detailed comparison between
QCD and SYM in the pertubative domain of both theories. Here we just
quote some of their results relevant for our discussion.

Firstly, notice that the differential photoemission rate can be
recast into the emission rate per unit volume as a function of the
photon momentum
\be
\frac{d\Gamma_\gamma}{dk} = \frac{\alpha_{\small \textrm{em}}}{\pi}
k \, \eta^{\mu\nu} \, C^<_{\mu\nu}(K) \, .
\ee
At low frequencies the trace of the spectral function approaches a
constant which is proportional to the electrical conductivity of the
plasma:
\be
\frac{d\Gamma_\gamma}{dk} = \frac{\sigma T}{\pi^2} k  \, .
\label{LowFreqPhotoemission}
\ee
In figure 5 we show the photoemission rates per unit volume per unit
time, divided by $\alpha_{\small \textrm{em}} (N^2-1) T^3$, both in
the non-perturbative and perturbative regimes. The non-pertubative
case is the same as before in figure 3 in the previous section, and
we repeat it here for comparison with the weakly coupled regime of
the SYM theory.

It is very interesting the fact that for weakly coupled ${\cal
{N}}=4$ supersymmetric Yang-Mills plasma the hydrodynamical regime
in which Eq.(\ref{LowFreqPhotoemission}) holds is as narrow as $k/T
\leq \lambda^2$, with $\lambda$ small. Its slope $\sigma T/\pi^2$ is
parametrically large and the photoemission rate has a maximum. For
larger photon momentum we take the expression from
\cite{CaronHuot:2006te}
\be
\frac{d\Gamma_\gamma}{dk} = \frac{(N^2-1) \alpha_{\small
\textrm{em}}}{4 \pi^2} k \, n_f(k) \, m_\infty^2 \,
[\ln(T/m_\infty)+C_{\small \textrm{tot}}(k/T)] \, .
\ee
being $n_f(k)$ the fermion statistical factor (with a minus sign),
while the thermal correction to the hard fermion propagation in the
medium is given by $m_\infty^2=\lambda T^2$. The integral called
$C_{\textrm{tot}}(k/T)$ was numerically solved in
\cite{CaronHuot:2006te} and also can be written in the following
form:
\be
C_{\small \textrm{tot}}(k/T)= \frac{1}{2} \ln(2k/T) + C_{2
\longleftrightarrow 2}(k/T)+C_{\small \textrm{brem}}(k/T)+C_{\small
\textrm{pair}}(k/T) \, . \label{W1}
\ee
The numerical results from \cite{CaronHuot:2006te} are reproduced
quite accurately by the expressions:
\ba
C_{2 \longleftrightarrow 2}(k/T) &\simeq& 2.01 \, T/k - 0.158 -
0.615 \, e^{-0.187 k/T} \, , \nonumber \\
C_{\small \textrm{brem}}(k/T)+C_{\small \textrm{pair}}(k/T) &\simeq&
0.954 \, (T/k)^{3/2} \, \ln(2.36 + T/k) + 0.069 + 0.0289 \, k/T \, , \nonumber \\
&&
\label{W2}
\ea
which hold in the range $0.2<k/T<20$. In figure 5 we have also shown
a small-dashed stiff curve which corresponds to the value of
$\lambda=0.2$ and a long-dashed curve for  $\lambda=0.5$, both
obtained using the combined expressions Eq.(\ref{W1}) and
Eq.(\ref{W2}). For weak coupling, it can be seen that the
photoemission rates decrease as the coupling decreases, except for
very low frequencies, where they are very much enhanced compared to
the strongly coupled regime. On the other hand, for large photon
momentum the strongly coupled plasma displays larger
photo-production rate. This effect is due to the suppression which
is produced by the $m_\infty$ factor, proportional to the 't Hooft
coupling, in the weakly coupled expression.
\begin{figure}
\begin{center}
\epsfig{file=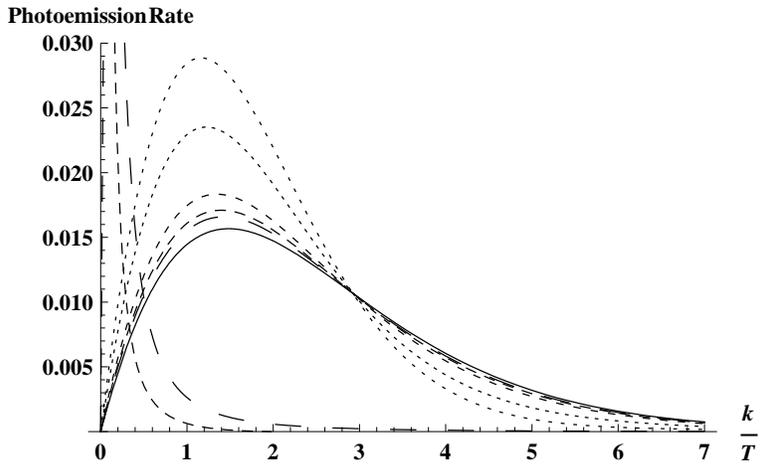, width=10cm}
{\caption{\small Photoemission spectrum for different values of
$\lambda$, as a function of the light-like momentum of the emitted
photon divided by the equilibrium temperature, $k/T$. Notice that in
fact we show the curves for $d\Gamma_\gamma/dk$ divided by
$\alpha_{\small \textrm{em}} (N^2-1) T^3$. Solid, long-dashed,
dashed, small-dashed, tiny-dashed, and dotted lines correspond to
decreasing values of $\lambda=$ very large (in fact it is the
analytical expression from supergravity with no string theory
corrections), to $200, 150, 100, 50 $ and 35, respectively. On the
other hand, we show two additional curves corresponding to the
weakly coupled SYM obtained in \cite{CaronHuot:2006te}:  a
small-dashed line and a long-dashed line, which represent the
perturbative SYM plasma for $\lambda=0.2$ and $0.5$, respectively.}}
\label{Scpectral}
\end{center}
\end{figure}
%

%%%%%%%%%%%%%%%%%%%%%%%%%%%%%%%%%%%%%%%%%%%%%%%%%%%%%%%%%%%%%%%%
\section{General discussion and conclusions}
%%%%%%%%%%%%%%%%%%%%%%%%%%%%%%%%%%%%%%%%%%%%%%%%%%%%%%%%%%%%%%%%

In this work we have investigated 't Hooft coupling corrections to
the photoemission rate of a strongly-coupled ${\cal {N}}=4$
supersymmetric Yang-Mills plasma by the means of the gauge/string
duality. We consider the full ${\cal {O}}(\alpha'^3)$ type IIB
string theory corrections to the supergravity action. The behaviour
of the photoemission rates depend upon the values of the 't Hooft
coupling. Their slopes at zero momentum give the electrical
conductivity as a function of the 't Hooft coupling. Beyond the
hydrodynamical regime of the plasma, as discussed before, the peak
of the photoemission is enhanced by the corrections, and the
momentum of maximal emission shifts towards the infrared, taking the
corrected curves closer to the weakly coupled result. Simple
numerical analysis on the light-like spectral function yields that
the maximal rate is given by
\begin{equation}
\left. \frac{d\Gamma_\gamma}{dk} \right\vert_{max} \simeq
0.0156695\left( 1 +\left[ 1115.3 - \frac{265}{8}\right]\gamma
\right) + {\cal {O}}(\gamma^2) \, ,
\end{equation}
in units of $\alpha_{\small \textrm{em}} N^2 T^3$, where we have
made explicit the factor $-265/8 \gamma$ coming from the overall
normalization of the action. For the peak displacement we estimate
the position of the peak $k_{max}$ as
\begin{equation}
k_{max} \simeq 1.48469\left( 1 - 842.425 \gamma  \right) T + {\cal
{O}}(\gamma^2) \, ,
\end{equation}
which turns out to be independent of the overall normalization of
the action, making it an excellent candidate for comparing disparate
gauge theories. Furthermore, we obtain the total number of photons
emitted, given by the area under the curves in figure 3. This is
enhanced by a factor
\begin{equation}
\frac{N_{\small\textrm{total}}(\gamma)}{N_{\small\textrm{total}}(0)}
\simeq 1 + \left[ 461.941 - \frac{265}{8} \right] \gamma + {\cal
{O}}(\gamma^2) \, ,
\end{equation}
due to the fact that the weakly-coupled theory dominates in the
infrared, where Bose-suppression (due to $n_b(k)$) is small.

These features are expected from perturbative quantum field theory
calculations in the weakly coupled ${\cal {N}}=4$ supersymmetric
Yang-Mills plasma and from the supergravity calculation of the large
$N$ strongly coupled theory \cite{CaronHuot:2006te}. There is a
($\lambda$-independent) crossover point around $k/T \sim 2.92$,
where the corrected curves dip below the $\lambda \rightarrow
\infty$ result. However, at much higher momentum, all these curves
cross upwards the extreme strongly coupled one, leading to that the
asymptotic values of the $\lambda$-corrected curves for large $k/T$
are given by $(1+5\gamma)$ times the infinite coupling result, as in
Eq.(\ref{chi-limit}). Although the range of momentum of figure 3
does not extend to cover this asymptotic behaviour, we show this
behaviour in figure 4. This means that the finite-$\lambda$
corrections enhance the photoemission rate in the deep ultraviolet
domain, contrary to the expectations of \cite{CaronHuot:2006te}.
Obviously, we are not guaranteed that the weakly-coupled result
should be approached by strongly-coupled corrections computed in
perturbation theory, especially not for a situation where the
functional dependence on momenta is expected to be different, so we
are not unduly concerned by this apparent discrepancy
\cite{Hassanain:2011ce}. It would be very revealing to understand
these cross-over points, as well as their scaling with $\lambda$.

The overall picture which emerges from our calculations is the
following: the peaks of the photoemission rates are displaced
towards low frequencies as $\lambda$ decreases. It implies that soft
photons are produced more efficiently at weaker values of the 't
Hooft coupling, in the strong coupling regime. On the other hand,
for harder photons the emission is dominated by stronger values of
the coupling. But then, due to the effect described in the previous
paragraph, the t' Hooft coupling corrected curves dominate for much
larger values of the ratio $k/T$. Still there is a notorious
separation of the behaviour at the weak and strong coupling regimes:
for the asymptotic behaviour of the weak coupling regime the fall
off goes proportional to $k^{3/2} e^{-k/T}$, while for all the
strongly coupled curves it falls down as $k^{5/2} e^{-k/T}$.

Let us consider very briefly what happens if we consider finite $N$
corrections. Using the modular form introduced in subsection 4.1,
since it factorises out from the Green-Stahn action
(\ref{green-stahn}) we obtain the corrections to the plasma
conductivity (recall that this is proportional to the slope of the
photoemission rate at zero frequency)
\be
\sigma = \sigma_0 + \frac{14993}{72} \, \sigma_0 \, \left(
\frac{\zeta(3)}{\lambda^{3/2}} + \frac{\lambda^{1/2}}{48 N^2} +
\frac{e^{-8 \pi^2 N/\lambda}}{2 \pi^{1/2} N^{3/2}} \right) \, .
\label{cond}
\ee
where as we have seen $\sigma_0 = e^2 T N^2/16 \pi$. At this point
one may wonder whether this result can be compared with those
obtained from lattice QCD. Obviously, any statement in the context
of the present work has to be considered with several caveats,
coming from differences between QCD and the ${\cal {N}}=4$ SYM
theory. Said that, it is possible to make contact with lattice QCD
at some extent. We must take into account that in lattice QCD
calculation $N=3$ typically, and there are other differences with
respect to the planar limit of the SYM plasma. In a recent
estimation of the electrical conductivity it was found $\sigma
\simeq 0.4 e^2 T$, above $T_c$ of quenched lattice QCD
\cite{Aarts:2007wj}. A more recent calculation \cite{Ding:2010ga}
shows that $1/3 e^2 T \leq \sigma \leq e^2 T$ from the vector
current correlation function from lattice computations at
temperatures about 1.5 to 2 $T_c$, the values of $\alpha_s =
g_{YM}^2/4 \pi$ are between 0.3 and 0.4, where these values were
obtained by matching the Debye mass screening in QCD and in ${\cal
{N}}=4$ SYM at finite $T$. If we use the parametrization
$\sigma=\rho e^2 T$ and extract $\rho$ from our equation
(\ref{cond}), using naively $N=3$, and evaluating the electrical
conductivity for $\lambda=11.3$, 15.08 and 6$\pi$, which lead to
$\alpha_s=$0.3, 0.4 and 0.5, respectively, we obtain $\rho=$1.64,
1.28 and 1.101, respectively. So, we can see how close is the lowest
value 1.101 to the upper value of the conductivity obtained from
lattice QCD in \cite{Ding:2010ga}.

The studies and results of photoemission and electrical conductivity
presented in this work exclusively concern the ${\cal {N}}=4$ SYM
plasma in thermal equilibrium. Beyond thermal equilibrium it is
possible to carry out very interesting computations such as the one
presented in \cite{Steineder:2012at}, where the authors consider
production of prompt photons from an out-of-equilibrium ${\cal
{N}}=4$ SYM plasma, including ${\cal {O}}(\alpha'^3)$ string theory
corrections. The work of \cite{Steineder:2012at} merges the
formalism introduced in \cite{Baier:2012at} about the
photo-production from a out-of-equilibrium strongly coupled plasma
with the one corresponding to ${\cal {O}}(\alpha'^3)$ string theory
corrections to photo-production \cite{Hassanain:2011ce}. The AdS/CFT
description of dilepton production from an out-of-equilibrium plasma
has also been considered \cite{Baier:2012tc}. Papers
\cite{Steineder:2012at,Baier:2012at,Baier:2012tc} are based on a
model of holographic thermalisation which involves the gravitational
collapse of a thin shell in $AdS_5$ in a quasi-static approximation
as in \cite{Danielsson:1999fa,Lin:2008rw}. On the other hand, one
can also consider other approach to holographic thermalisation with
a dynamical shell as in
\cite{AbajoArrastia:2010yt,Balasubramanian:2011ur}. Based on
\cite{AbajoArrastia:2010yt,Balasubramanian:2011ur}, thermal and
electromagnetic quenches for the specific case of $AdS_4$ have been
studied \cite{Albash:2010mv}. More recently, a systematic study of
holographic thermalisation from a collapsing shell of charged
pressureless dust has been considered in \cite{Galante:2012pv},
which allows one to consider a chemical potential effect on the
thermalisation time scale of the plasma.

There are very interesting directions for further extensions of the
work presented here. One is to look for similar strong coupling
corrections to the electrical diffusion constant, with the idea of
discussing about the possibility of a universal bound for electrical
charge transport coefficients \cite{Kovtun:2003wp,Ritz:2008kh}.
Also, effects of leading 't Hooft coupling corrections to dilepton
production from a strongly coupled plasma would be interesting to
study in this context. Another interesting situation is the boost
invariant plasma. In that case corrections coming from $F_5$ could
have modified the ratio of shear viscosity over entropy density,
however in \cite{Myers:2008yi} it was concluded that corrections
from $F_5$ do not modify the results in this case. However, likely
the situation would be different if one refers to electric charge
transport properties. In addition, the inclusion of an $R$-charged
black hole allows to study the plasma at finite chemical potential.
In this case, as Paulos has shown \cite{Paulos:2008tn}, the EOM are
modified by the $F_5$ corrections. However, while the analysis using
only $C^4$-term is prohibitively complicated, Paulos has shown that
the inclusion of its supersymmetric completion gives a simple
result. It would be interesting to see if something similar happens
for properties related to electric charge in the same background.

It would also be very interesting to carry out similar calculation
as for the corrections to electrical conductivity and photoemission
rates for other backgrounds with deformations from the conformal
ones \cite{Polchinski:2000uf,Pilch:2000fu} and with no conformal
symmetry \cite{Klebanov:2000hb,Maldacena:2000yy}. It would also be
very interesting to be able to compute higher-order corrections to
the photoemission rate for plasmas with fundamental quarks. In that
case there are at least two different and very interesting
possibilities. One is the case for a D3D7 plasma which basically
consists of the embedding of D7 branes in the background of a large
number of D3 branes \cite{Kruczenski:2003be} at finite temperature.
There are two possible embeddings for this: the Minkowski and the
black hole ones, and there is a Hawking-Page transition which is
associated with glueball to deconfined gluons transition and at
higher temperature there is a second transition temperature at which
mesons melt down giving a QGP \cite{Mateos:2007vn}. Another very
different system which actually is closer to QCD than the D3D7 brane
system is the D4D8-anti D8-brane system proposed by Sakai and
Sugimoto \cite{Sakai:2004cn}, which can also be heated up in order
to obtain a QGP \cite{Peeters:2006iu}. $\alpha'$ corrections to
D-brane solutions have been investigated in \cite{deHaro:2003zd}.

~

~

%%%%%%%%%%%%%%%%%%%%%%%%%%%%%%%%%%%%%%%%%%%%%%
\centerline{\large{\bf Acknowledgments}}
%%%%%%%%%%%%%%%%%%%%%%%%%%%%%%%%%%%%%%%%%%%%%%

~

Authors thank Gert Aarts, Alex Buchel, Carlos N\'u\~nez, Miguel
Paulos, Jorge Russo, Dominik Steineder, Stefan Stricker and Aleksi
Vuorinen for valuable correspondence, comments and discussions. The
work of M.S. has been partially supported by the CONICET, the
ANPCyT-FONCyT Grant PICT-2007-00849, and the CONICET Grant
PIP-2010-0396.

\newpage

\subsection*{Appendix: The magnetic part of $F_5$ for vector fluctuations}

~

{\bf The vector fluctuation $F_{ux}$} \\

Let us first consider the vector fluctuation $F_{ux}$. The magnetic
part is a ten-dimensional Hodge dual of the electric one. In
components we can write
\be
F_{(m)u x y_{\nu_1} y_{\nu_2} y_{\nu_3}}^f = \sqrt{|\det G_{10}|} \,
F_{(e)y_i y_j tyz}^f \, G^{y_i y_i} \, G^{y_j y_j} \, G^{tt} \,
G^{yy} \, G^{zz} \, \epsilon_{y_i y_j tyzxu \nu_1 \nu_2 \nu_3} \, .
\ee
Note that this component as a form reads:
\be
F_{(m)}^f = \frac{\sqrt{|\det G_{10}|}}{5! 5!} \, F_{(e)y_i y_j
tyz}^f \, G^{y_i y_i} \, G^{y_j y_j} \, G^{tt} \, G^{yy} \, G^{zz}
\, \epsilon_{y_i y_j t y z x u \nu_1 \nu_2 \nu_3} \, du \wedge dx
\wedge dy^{\nu_1} \wedge dy^{\nu_2} \wedge dy^{\nu_3}  \, ,
\ee
$\epsilon_{y_i y_j t y z x u \nu_1 \nu_2 \nu_3}$ determines the sign
of each piece of the magnetic components as follows.

So, let us label the magnetic components by the indices of the
vector fluctuation in the metric, {\it i.e.} ${u x}$ in this case we
have
\ba
F_{(m)ux}^f &=& \sqrt{|\det g_{AdS}|} \, F_{(e)ux}^f \,  G^{tt} \,
G^{yy} \, G^{zz} \, (m_{13}^{ux} + m_{14}^{ux} + m_{15}^{ux} +
m_{24}^{ux} + m_{25}^{ux} ) \, .
\ea
Then
\ba
m_{13}^{ux} &=& \epsilon_{13tyzxu245} \, \sqrt{\det S^5} \, b_{13}
\, G^{y_1 y_1} \, G^{y_3 y_3} = - \sqrt{\det S^5} \, b_{13} \,
G^{y_1 y_1} \, G^{y_3 y_3} \, ,\nonumber
\\
m_{14}^{ux} &=& \epsilon_{14tyzxu235} \, \sqrt{\det S^5} \, b_{14}
\, G^{y_1 y_1} \, G^{y_4 y_4} = + \sqrt{\det S^5} \, b_{14}  G^{y_1
y_1} \, G^{y_4 y_4} \, , \nonumber
\\
m_{15}^{ux} &=& \epsilon_{15tyzxu234} \, \sqrt{\det S^5} \, b_{15}
\, G^{y_1 y_1} \, G^{y_5 y_5}= - \sqrt{\det S^5} \, b_{15} \, G^{y_1
y_1} \, G^{y_5 y_5} \, , \nonumber
\\
m_{24}^{ux} &=& \epsilon_{24tyzxu135} \, \sqrt{\det S^5} \, b_{24}
\, G^{y_2 y_2} \, G^{y_4 y_4} = - \sqrt{\det S^5} \, b_{24} \,
G^{y_2 y_2} \, G^{y_4 y_4} \, , \nonumber
\\
m_{25}^{ux} &=& \epsilon_{25tyzxu134} \, \sqrt{\det S^5} \, b_{25}
\, G^{y_2 y_2} \, G^{y_5 y_5}= + \sqrt{\det S^5} \, b_{25} \, G^{y_2
y_2} \, G^{y_5 y_5}  \, . \nonumber
\\
\ea

~

{\bf The vector fluctuation $F_{tx}$} \\

As before, the magnetic part is a 10d Hodge dual of the electric
one. In components we have
\be
F_{(m)t x y_{\nu_1} y_{\nu_2} y_{\nu_3}}^f = \sqrt{|\det G_{10}|} \,
F_{(e)y_i y_j y z u}^f \, G^{y_i y_i} \, G^{y_j y_j} \, G^{uu} \,
G^{yy} \, G^{zz} \, \epsilon_{y_i y_j y z u t x \nu_1 \nu_2 \nu_3}
\, ,
\ee
As a form it reads:
\be
F_{(m)}^f = \frac{\sqrt{|\det G_{10}|}}{5! 5!} \, F_{(e)y_i y_j y z
u}^f \, G^{y_i y_i} \, G^{y_j y_j} \, G^{uu} \, G^{yy} \, G^{zz} \,
\epsilon_{y_i y_j y z u t x \nu_1 \nu_2 \nu_3} \, dt \wedge dx
\wedge dy^{\nu_1} \wedge dy^{\nu_2} \wedge dy^{\nu_3}  \, .
\ee
Again, let us label the magnetic components by the indices of the
vector fluctuation in the metric, {\it i.e.} ${t x}$
\ba
F_{(m)tx}^f &=& \sqrt{|\det g_{AdS}|} \, F_{(e)tx}^f \,  G^{uu} \,
G^{yy} \, G^{zz} \, (m_{13}^{tx} + m_{14}^{tx} + m_{15}^{tx}  +
m_{24}^{tx}  + m_{25}^{tx} ) \, .
\ea
Then we have,
\ba
m_{13}^{tx} &=& \epsilon_{13yzutx245} \, \sqrt{\det S^5} \, b_{13}
\, G^{y_1 y_1} \, G^{y_3 y_3} = - \, \sqrt{\det S^5} \, b_{13} \,
G^{y_1 y_1} \, G^{y_3 y_3} \, , \nonumber
\\
m_{14}^{tx} &=& \epsilon_{14yzutx235} \, \sqrt{\det S^5} \, b_{14}
\, G^{y_1 y_1} \, G^{y_4 y_4} = + \, \sqrt{\det S^5} \, b_{14} \,
G^{y_1 y_1} \, G^{y_4 y_4} \, , \nonumber
\\
m_{15}^{tx} &=& \epsilon_{15yzutx234} \, \sqrt{\det S^5} \, b_{15}
\, G^{y_1 y_1} \, G^{y_5 y_5} = - \, \sqrt{\det S^5} \, b_{15} \,
G^{y_1 y_1} \, G^{y_5 y_5} \, , \nonumber
\\
m_{24}^{tx} &=& \epsilon_{24yzutx135} \, \sqrt{\det S^5} \, b_{24}
\, G^{y_2 y_2} \, G^{y_4 y_4} = - \, \sqrt{\det S^5} \, b_{24} \,
G^{y_2 y_2} \, G^{y_4 y_4} \, , \nonumber
\\
m_{25}^{tx} &=& \epsilon_{25yzutx134} \, \sqrt{\det S^5} \, b_{25}
\, G^{y_2 y_2} \, G^{y_5 y_5} = + \, \sqrt{\det S^5} \, b_{25} \,
G^{y_2 y_2} \, G^{y_5 y_5} \, . \nonumber
\\
\ea

~

{\bf The vector fluctuation $F_{zx}$} \\

Similarly with the two previous cases the ten-dimensional Hodge dual
of electrical part of $F_5$ induced by the vector fluctuation
$F_{zx}$ in components is given by
\be
F_{(m)z x y_{\nu_1} y_{\nu_2} y_{\nu_3}}^f = \sqrt{|\det G_{10}|} \,
F_{(e)y_i y_j t y u}^f \, G^{y_i y_i} \, G^{y_j y_j} \, G^{tt} \,
G^{yy} \, G^{uu} \, \epsilon_{y_i y_j t y u x z \nu_1 \nu_2 \nu_3}
\, .
\ee
This component as a form reads:
\be
F_{(m)}^f = \frac{\sqrt{|\det G_{10}|}}{5! 5!} \, F_{(e)y_i y_j t y
u}^f \, G^{y_i y_i} \, G^{y_j y_j} \, G^{tt} \, G^{yy} \, G^{uu} \,
\epsilon_{y_i y_j t y u x z \nu_1 \nu_2 \nu_3} \, dx \wedge dz
\wedge dy^{\nu_1} \wedge dy^{\nu_2} \wedge dy^{\nu_3}  \, .
\ee
Now, let us label the magnetic components by the indices of the
vector fluctuation in the metric, {\it i.e.} ${z x}$
\ba
F_{(m)zx}^f &=& \sqrt{|\det g_{AdS}|} \, F_{(e)ux}^f \,  G^{tt} \,
G^{yy} \, G^{uu} \, (m_{13}^{zx} + m_{14}^{zx} + m_{15}^{zx} +
m_{24}^{zx} + m_{25}^{zx} ) \ .
\ea
Then
\ba
m_{13}^{zx} &=& \epsilon_{13tyuxz245} \, \sqrt{\det S^5} \, b_{13}
\, G^{y_1 y_1} \, G^{y_3 y_3}= - \, \sqrt{\det S^5} \, b_{13} \,
G^{y_1 y_1} \, G^{y_3 y_3} \, (-1) \, , \nonumber
\\
m_{14}^{zx} &=& \epsilon_{14tyuxz235} \, \sqrt{\det S^5} \, b_{14}
\, G^{y_1 y_1} \, G^{y_4 y_4}= + \, \sqrt{\det S^5} \, b_{14} \,
G^{y_1 y_1} \, G^{y_4 y_4} \, (-1) \, , \nonumber
\\
m_{15}^{zx} &=& \epsilon_{15tyuxz234} \, \sqrt{\det S^5} \, b_{15}
\, G^{y_1 y_1} \, G^{y_5 y_5}= - \, \sqrt{\det S^5} \, b_{15} \,
G^{y_1 y_1} \, G^{y_5 y_5} \, (-1) \, , \nonumber
\\
m_{24}^{zx} &=& \epsilon_{24tyuxz135} \, \sqrt{\det S^5} \, b_{24}
\, G^{y_2 y_2} \, G^{y_4 y_4}= - \, \sqrt{\det S^5} \, b_{24} \,
G^{y_2 y_2} \, G^{y_4 y_4} \, (-1) \, , \nonumber
\\
m_{25}^{zx} &=& \epsilon_{25tyuxz134} \, \sqrt{\det S^5} \, b_{25}
\, G^{y_2 y_2} \, G^{y_5 y_5}= + \, \sqrt{\det S^5} \, b_{25} \,
G^{y_2 y_2} \, G^{y_5 y_5} \, (-1) \, . \nonumber
\\
\ea

\end{document}